\documentclass[12pt]{article}
\usepackage{amsfonts}
\usepackage{epsfig,amsmath,latexsym,amssymb}
\usepackage{amscd}
\usepackage{multirow}
\usepackage{graphicx}
\usepackage{geometry}
\usepackage{lscape}
\usepackage{amsmath}
\usepackage{amssymb}
\usepackage{bm}%
\providecommand{\U}[1]{\protect\rule{.1in}{.1in}}
\def\R{{\mathbb R}}

\newcommand{\Remm}[1]{}

\newtheorem{model ass}[theo]{Model Assumptions}

\numberwithin{equation}{section}

\begin{document}

\title{Model uncertainty in claims reserving within Tweedie's compound Poisson models}

\author{Gareth W.~Peters \quad Pavel V.~Shevchenko \quad Mario
V.~W\"{u}thrich}

\date{\footnotesize{First version: 5 December 2007 \\This version: 13 October 2008}}

\maketitle

\begin{center}
This is a preprint of an article to appear in \\ASTIN Bulletin
39(1), pp.1-33, 2009.
\end{center}


\begin{abstract}
\noindent In this paper we examine the claims reserving problem
using Tweedie's compound Poisson model. We develop the maximum
likelihood and Bayesian Markov chain Monte Carlo simulation
approaches to fit the model and then compare the estimated models
under different scenarios. The key point we demonstrate relates to
the comparison of reserving quantities with and without model
uncertainty incorporated into the prediction. We consider both the
model selection problem and the model averaging solutions for the
predicted reserves. As a part of this process we also consider the
sub problem of variable selection to obtain a parsimonious
representation of the model being fitted.

\vspace{1cm}

\end{abstract}
\pagebreak

\section{Claims reserving}

Setting appropriate claims reserves to meet future claims payment
cash flows is one of the main tasks of non-life insurance
actuaries. There is a wide range of models, methods and algorithms
used to set appropriate claims reserves. Among the most popular
methods there is the chain-ladder method, the Bornhuetter-Ferguson
method and the generalized linear model methods. For an overview,
see W\"uthrich and Merz (2008) and England and Verrall (2002).

Setting claims reserves includes two tasks: estimate the mean of
future payments and quantify the uncertainty in this prediction for
future payments. Typically, quantifying the uncertainty includes two
terms, namely the so-called process variance and the (parameter)
estimation
error. The process variance reflects that we predict random
variables, i.e. it describes the pure process uncertainty. The
estimation error reflects that the true model parameters need to be
estimated and hence there is an uncertainty in the reliability of
these estimates. In this paper, in addition to these two terms, we
consider a third source of error/uncertainty, namely, we analyze the
fact that we could have chosen the wrong model. That is, we select a
family of claims reserving models and quantify the uncertainty coming
from a possibly wrong model choice within this family of models.

Such an analysis is especially important when answering solvency
questions. A poor model choice may result in a severe shortfall in
the balance sheet of an insurance company, which requires under a
risk-adjusted solvency regime an adequate risk capital charge. We
analyze typical sizes of such risk capital charges within the family
of Tweedie's compound Poisson models\textbf{, }see Tweedie (1984),
Smyth and J{\o}rgensen (2002) and W\"{u}thrich (2003).

Assume that $Y_{i,j}$ are incremental claims payments with indices
$i,j\in\left\{  0,\ldots,I\right\}  ,$ where $i$ denotes the
accident year and $j$ denotes the development year. At time $I$, we
have observations
\begin{equation}
\mathcal{D}_{I}=\left\{  Y_{i,j};~i+j\leq I\right\}
\end{equation}
and for claims reserving at time $I$ we need to predict the future payments
\begin{equation}
\mathcal{D}_{I}^{c}=\left\{  Y_{i,j};~i+j>I,i\leq I\right\}  ,
\end{equation}
see Table \ref{tab1}. Hence, the outstanding claims payment at time $I$ is
given by
\begin{equation}
R=\sum_{i=1}^{I}R_{i}=\sum_{i+j>I}Y_{i,j}. \label{R}%
\end{equation}
Its conditional expectation at time $I$ is given by
\begin{equation}
E\left[  \left.  R\right\vert \mathcal{D}_{I}\right]  =\sum_{i=1}^{I}E\left[
\left.  R_{i}\right\vert \mathcal{D}_{I}\right]  =\sum_{i+j>I}E\left[  \left.
Y_{i,j}\right\vert \mathcal{D}_{I}\right]  .
\end{equation}
Hereafter, the summation $i+j>I$ is for $i\leq I$. Therefore, we
need to predict $R$ and to estimate $E\left[  \left.  R\right\vert
\mathcal{D}_{I}\right]  $. Assume that $\widehat{R}$ is an
appropriate $\mathcal{D}_{I}$-measurable predictor  for $R$ and
$\mathcal{D}_{I}$-measurable estimator for $E\left[  \left.
R\right\vert \mathcal{D}_{I}\right] $. Then, $\widehat{R}$ is used to predict the future payments and is the
amount that is put aside in the balance sheet of the insurance
company for these payments.

Prediction uncertainty is then often studied with the help of the
(conditional) mean square error of prediction (MSEP) which is
defined by
\begin{equation}
\mathrm{msep}_{R|\mathcal{D}_{I}}\left(  \widehat{R}\right)  =E\left[  \left.
\left(  R-\widehat{R}\right)  ^{2}\right\vert \mathcal{D}_{I}\right]  .
\end{equation}
If $\widehat{R}$ is $\mathcal{D}_{I}$-measurable, the conditional
MSEP can easily be decoupled as follows\textbf{, }see W\"uthrich
and Merz (2008), section 3.1:
\begin{align}
\mathrm{msep}_{R|\mathcal{D}_{I}}\left(  \widehat{R}\right)   &
=\mathrm{Var}\left(  \left.  R\right\vert \mathcal{D}_{I}\right)  +\left(
E\left.  \left[  R\right\vert \mathcal{D}_{I}\right]  -\widehat{R}\right)
^{2}\label{vardecomp}\\
&  =\text{ process variance }+\text{ estimation error.}\nonumber
\end{align}
It is clear that the consistent estimator $\widehat{R}$ which
minimizes the conditional MSEP is given by $\widehat{R}=E\left.
\left[ R\right\vert \mathcal{D}_{I}\right]  $ and is used,
hereafter, as the "best estimate" for reserves. Assuming the model
is parameterized by the parameter vector $\bm{\theta}$,
$\mathrm{Var}\left(  \left.  R\right\vert \mathcal{D} _{I}\right)$
can be decomposed as
\begin{align}
\mathrm{Var}\left(  \left.  R\right\vert \mathcal{D}_{I}\right)   &
=E\left[\left. \mathrm{Var}\left(  \left.  R\right\vert \bm{\theta},
\mathcal{D}_{I}\right)\right\vert \mathcal{D}_{I}\right]
+\mathrm{Var}\left(\left. E\left[ \left. R\right\vert \bm{\theta},
\mathcal{D}_{I}\right] \right\vert \mathcal{D}_{I}\right)
\label{decompvar}\\
&  =\text{average process variance }+\text{ parameter estimation
error}.\nonumber
\end{align}
These are the two terms that are usually studied when quantifying
prediction uncertainties in a Bayesian context, where
the unknown parameters $\bm{\theta }$ are modelled stochastically.
That is, we obtain in the Bayesian context
a similar decomposition as in the frequentist estimation
(\ref{vardecomp}).
In the frequentist approach, the second term in (\ref{vardecomp}) is
often estimated by $\mathrm{Var}( \widehat{R})$, see for example
section 6.4.3 in W\"uthrich and Merz (2008).

As discussed in Cairns (2000), in full generality one could consider
several sources of model uncertainty, however unlike Cairns (2000)
we focus on a specific class of models. We consider the setting
discussed in Bernardo and Smith (1994) termed M Complete modelling.
In such a setting the premise is that one considers a set of models
in which the "truth" exists but is unknown \textit{a priori}. In
this setting we demonstrate the risk associated with the model
uncertainty which we analyze jointly as a decomposition into two
main parts. The first involves the uncertainty in the
parameterization of the model, this is a variable selection problem
within a nested model structure in the same vein as discussed in
Cairns (2000). It relates to finding a trade-off between parsimony
and accuracy in the estimation.\ The second source of model
uncertainty that we study involves the choice of a parameter which
determines membership from a spectrum of possible models within the
Tweedie's compound Poisson family of models. We restrict the
analysis to  Tweedie's compound Poisson models and justify this by
assuming we are working in the M Complete setting. If we relaxed
this assumption and therefore consider competing models not in this
family, then the analysis would be difficult to interpret and
analyze in the manner we develop in this paper. The second source of
model uncertainty will be considered under both a model selection
and a model averaging setting, given the first "variable selection"
uncertainty is resolved. As mentioned in Cairns (2000) achieving
such an analysis requires advanced simulation methodology. Note, in
future work we would also consider the M Open modeling framework of
Bernardo and Smith (1994) which relaxes the belief that the truth
lies in the set of models considered and hence introduces additional
uncertainty associated with the family of models considered. The
advanced sampling methodology required to study the M Open model
setting will be briefly discussed.

The paper is organised as follows. In section 2, we present
Tweedie's compound Poisson model and section 3 considers parameter
estimation in the model, using the maximum Likelihood and Bayesian
Markov chain Monte Carlo approaches for a real data set. Having
addressed the variable selection question in section 4, we then
analyze claims reserve estimation and model uncertainty in both a
frequentist and Bayesian setting in section 5. We finish with
conclusions from our findings.

\section{Tweedie's compound Poisson model}

We assume that $Y_{i,j}$ belongs to the family of Tweedie's compound
Poisson models. Below we provide three different parameterizations
for Tweedie's compound Poisson models,  for rigorous derivations we
refer to J{\o}rgensen and de Souza (1994), Smyth and J{\o}rgensen
(2002) and W\"{u}thrich (2003).

\begin{model ass}
[1st Representation]\label{model ass1} \textrm{We assume that
$Y_{i,j}$ are independent for $i,j\in\left\{  0,\ldots,I\right\}$
and have a compound Poisson distribution
\begin{equation}
Y_{i,j}=1_{\{N_{i,j}>0\}}~\sum_{k=1}^{N_{i,j}}X_{i,j}^{(k)}
,\label{Tweedie Model}
\end{equation}
in which (a) $N_{i,j}$ and $X_{i,j}^{(k)}$ are independent for all
$k$, (b) $N_{i,j}$ is Poisson distributed with parameter
$\lambda_{i,j}$; (c) $X_{i,j}^{(k)}$ are independent gamma
severities with the mean $\tau_{i,j}>0$ and the shape parameter
$\gamma>0$. Hereafter, we denote }$1_{\{\}}$ as an indicator
function.
\end{model ass}

~

\textbf{2nd Representation. }The random variable $Y_{i,j}$ given in
\eqref{Tweedie Model} belongs to the family of Tweedie's compound
Poisson models, see Tweedie (1984). The distribution of $Y_{i,j}$
can be reparameterized in such a way that it takes a form of the
exponential dispersion family, see e.g. formula (3.5) and Appendix A
in W\"{u}thrich (2003):
\newline$Y_{i,j}$ has a probability weight at $0$ given by
\begin{equation}
P\left[  Y_{i,j}=0\right]  =P\left[  N_{i,j}=0\right]  =\exp\left\{
-\phi_{i,j}^{-1}\kappa_{p}(\theta_{i,j})\right\}  \label{tweedie}
\end{equation}
and for $y>0$ the random variable $Y_{i,j}$ has continuous density
\begin{equation}
f_{\theta_{i,j}}(y;\phi_{i,j},p)=c(y;\phi_{i,j},p)\exp\left\{  \frac
{y~\theta_{i,j}-\kappa_{p}(\theta_{i,j})}{\phi_{i,j}}\right\}.
\end{equation}

Here $\theta_{i,j}<0$, $\phi_{i,j}>0$, the normalizing constant is
given by
\begin{equation}
c(y;\phi,p)=\sum_{r\geq1}\left(  \frac{(1/\phi)^{\gamma+1}y^{\gamma}
}{(p-1)^{\gamma}(2-p)}\right)  ^{r}\frac{1}{r!\Gamma(r\gamma)y}
\label{tweedie2}%
\end{equation}
and the cummulant generating function $\kappa_{p}(.)$ is given by%
\begin{equation}
\kappa_{p}(\theta)\overset{def.}{=}\frac{1}{2-p}\left[
(1-p)\theta\right] ^{\gamma},
\end{equation}
where $p\in\left(  1,2\right)  $ and $\gamma=\left(  2-p\right)  /\left(
1-p\right)  .$

The parameters, in terms of the 1st representation quantities, are:
\begin{align}
p=p(\gamma)  &  =\frac{\gamma+2}{\gamma+1}~\in~(1,2),\label{def p}\\
\phi_{i,j}  &  =\frac{\lambda_{i,j}^{1-p}\tau_{i,j}^{2-p}}{2-p}>0,\\
\theta_{i,j}  &  =\left(  \frac{1}{1-p}\right)  \left(  \mu_{i,j}\right)
^{\left(  1-p\right)  }<0,\label{params}\\
\mu_{i,j}  &  =\lambda_{i,j}\tau_{i,j}>0.
\end{align}

Then the mean and variance of $Y_{i,j}$ are given by
\begin{align}
E\left[  Y_{i,j}\right]   &  =\frac{\partial}{\partial\theta_{i,j}}\kappa
_{p}(\theta_{i,j})=\kappa_{p}^{\prime}(\theta_{i,j})=\left[  (1-p)\theta
_{i,j}\right]  ^{1/(1-p)}=\mu_{i,j},\\
\text{Var}\left(  Y_{i,j}\right)   &  =\phi_{i,j}\kappa_{p}^{\prime\prime
}(\theta_{i,j})=\phi_{i,j}~\mu_{i,j}^{p}.
\end{align}
That is, $Y_{i,j}$ has the mean $\mu_{i,j}$, dispersion $\phi_{i,j}$
and variance function with the variance parameter $p$. The extreme
cases $p\rightarrow1$ and $p\rightarrow2$ correspond to the
overdispersed Poisson and the gamma models, respectively. Hence, in
this spirit, Tweedie's compound Poisson model with $p\in\left(
1,2\right)  $ closes the gap between the Poisson and the gamma
models. Often in practice, $p$ is assumed to be known and fixed by
the modeller. The aim of this paper is to study \textit{Model
Uncertainty}, that is, we would like to study the sensitivity of the
claims reserves within this subfamily, i.e.~Tweedie's compound
Poisson models (which are now parameterized through $p$). This
answers model uncertainty questions within the family of Tweedie's
compound Poisson models. In this paper the restriction on
$p\in\left( 1,2\right)  $ is taken in the context of practical
application of these models to claims reserving, W\"{u}thrich (2003)
comments that the majority of claims reserving problems will be
captured under this assumption. However, in general, in the
exponential dispersion family $p$ can be outside of the $\left(
1,2\right)  $ range, e.g.  $p=0$ produces a Gaussian density and
$p=3$ leads to an inverse Gaussian model.

~

\textbf{3rd Representation.} Utilizing the above definitions, the
distribution of $Y_{i,j}$ can be rewritten in terms of $\mu_{i,j}$,
$p$ and $\phi_{i,j}$ as
\begin{equation}
P\left[  Y_{i,j}=0\right]  =P\left[  N_{i,j}=0\right]  =\exp\left\{
-\phi_{i,j}^{-1}\frac{\mu_{i,j}^{2-p}}{2-p}\right\}
\end{equation}
and for $y>0$%
\begin{equation}
f_{\mu_{i,j}}(y;\phi_{i,j},p)=c(y;\phi_{i,j},p)\exp\left\{  \phi_{i,j}%
^{-1}\left[  y~\frac{\mu_{i,j}^{1-p}}{1-p}-\frac{\mu_{i,j}^{2-p}}{2-p}\right]
\right\}  .
\end{equation}

\section{Parameter estimation}

Our goal is to estimate the parameters $\mu_{i,j}$, $p$ and $\phi_{i,j}$ based
on the observations $\mathcal{D}_{I}$. In order to estimate these parameters
we need to introduce additional structure in the form of a multiplicative model.

\begin{model ass}
\textrm{Assume that there exist exposures $\bm{\alpha}=\left(  \alpha
_{0},\ldots,\alpha_{I}\right)  $ and a development pattern
$\bm{\beta }=\left(  \beta_{0},\ldots,\beta_{I}\right)  $ such that we have
for all $i,j\in\{0,\ldots,I\}$
\begin{equation}
\mu_{i,j}=\alpha_{i}~\beta_{j}.
\end{equation}
Moreover, assume that $\phi_{i,j}=\phi$ and\
}$\alpha_{i}>0$,$~\beta_{j}
>0$\textrm{. }
\end{model ass}

In addition, we impose the normalizing condition $\alpha_{0}=1$, so
that the estimation problem is well-defined. That is we have $\left(
2I+3\right)  $ unknown parameters $p,\phi, \bm{\alpha,\beta}$ that
have to be estimated from the data $\mathcal{D}_{I}$. Next we
present the likelihood function for this model and then develop the
methodology for parameter estimation using the maximum likelihood
and Bayesian inference methods.

\subsection{Likelihood function}

Define the parameter vector $\bm{\theta}=\left(p,\phi,
\mathrm{\bm{\alpha}} {,}\mathrm{\bm{\beta }}\right)  $. Then the
likelihood function for $Y_{i,j}$, $i+j\leq I$, is given by
\begin{align}
L_{\mathcal{D}_{I}}(\bm{\theta})  &  =\prod_{i+j\leq
I}c(Y_{i,j};\phi ,p)\exp\left\{  \phi^{-1}\left[
Y_{i,j}\frac{(\alpha_{i}\beta_{j})^{1-p}
}{1-p}-\frac{(\alpha_{i}\beta_{j})^{2-p}}{2-p}\right]  \right\},
\label{LikelihoodFunction}
\end{align}
where we set $c(0;\phi ,p)=1$ for $Y_{i,j}=0$. The difficulty in the
evaluation of the likelihood function is the calculation of $c(y;\phi,p)$
which contains an infinite sum
\begin{equation}
c(y;\phi,p)=\sum_{r\geq1}\left(  \frac{(1/\phi)^{\gamma+1}y^{\gamma}
}{(p-1)^{\gamma}(2-p)}\right)
^{r}\frac{1}{r!\Gamma(r\gamma)y}=\frac{1} {y}\sum_{r\geq1}W_{r},
\end{equation}
where $\gamma=\gamma\left(  p\right)  =\left(  2-p\right)  /\left(
1-p\right)$. Tweedie (1984) identified this summation as Wright's
(1935) generalized Bessel function, which can not be expressed in
terms of more common Bessel functions. To evaluate this summation we
follow the approach of Dunn and Smyth (2005) which directly sums the
infinite series, including only terms which significantly contribute
to the summation. Consider the term
\[
\log W_{r}=r\log z-\log\Gamma\left(  1+r\right)  -\log\Gamma\left(  \gamma
r\right)  ,
\]
where
\[
z=\frac{(1/\phi)^{\gamma+1}y^{\gamma}}{\left(  p-1\right)  ^{\gamma}\left(
2-p\right)  }.
\]
Replacing the gamma functions using Stirling's approximation and approximating
$\gamma r$ by $\gamma r+1$ we get
\[
\log W_{r}\approx r\left\{  \log z+\left(  1+\gamma\right)  -\gamma\log
\gamma-\left(  1+\gamma\right)  \log r\right\}  -\log\left(  2\pi\right)
-\frac{1}{2}\log\gamma-\log r,
\]
which is also a reasonable approximation for small $r$. Treating $r$ as
continuous and taking the partial derivative w.r.t.~$r$ gives%
\[
\frac{\partial\log W_{r}}{\partial r}\approx\log z-\log r-\gamma\log\left(
\gamma r\right)  .
\]
Hence, the sequence $W_{r}$ is unimodal in $r.$ Solving $\partial
W_{r}/\partial r=0$, to find (approximately) the maximum of $W_{r}$, results
in the approximate maximum lying close to
\begin{equation}
R_{0}=R_{0}\left(  \phi,p\right)  =\frac{y^{2-p}}{\left(  2-p\right)  \phi}.
\end{equation}
This gives a surprisingly accurate approximation to the true maximum of
$W_{r}$, $r\in{\mathbb{N}}$. Finally, the aim is to find $R_{L}%
<R_{0}<R_{U}$ such that the following approximation is sufficiently accurate
for the use in the evaluation of the likelihood terms,%
\begin{equation}
c(y;\phi,p)\approx\widetilde{c}(y;\phi,p)=\frac{1}{y}\sum_{r=R_{L}}^{R_{U}
}W_{r}.
\end{equation}
The fact that $\partial\log W_{r}/\partial r$ is monotonic and
decreasing implies that $\log W_{r}$ is strictly convex in $r$ and
hence the terms in $W_{r}$ decay at a faster rate than geometric on
either side of $R_{0}$. Dunn and Smyth (2005) derive the following
bounds,
\begin{equation}
c(y;\phi,p)-\widetilde{c}(y;\phi,p)<W_{R_{L}-1}\frac{1-q_{L}^{R_{L}-1}
}{1-q_{L}}+W_{R_{U}+1}\frac{1}{1-q_{U}}%
\end{equation}
with
\begin{equation}
q_{L} =\left.  \exp\left(  \frac{\partial\log W_{r}}{\partial
r}\right) \right\vert _{r=R_{L}-1},\text{ \ \ } q_{U}=\left.
\exp\left( \frac{\partial\log W_{r}}{\partial r}\right) \right\vert
_{r=R_{U}+1}.
\end{equation}
These bounds are typically too conservative since the decay is much
faster than geometric. In practice, an adaptive approach balancing
accuracy and efficiency is to continue adding terms either side of
the maximum until
the lower and upper terms satisfy the double precision constraints $W_{R_{L}%
}\leqslant e^{-37}W_{R_{0}}$ (or $R_{L}=1$) and $W_{R_{U}}\leqslant
e^{-37}W_{R_{0}}$. When evaluating the summation for
$\widetilde{c}(y;\phi ,p)$, it was important to utilize the
following identity to perform the summation in the log scale to
avoid numerical overflow problems,
\[
\log\widetilde{c}(y;\phi,p)=-\log y+\log W_{R_{0}}+\log\left(  \sum
\limits_{r=R_{L}}^{R_{U}}\exp\left(  \log\left(  W_{R}\right)  -\log\left(
W_{R_{0}}\right)  \right)  \right)  .
\]

We made an additional observation when analyzing this model. For our
data set, as $p$ approaches $1$ (i.e. when \ the distribution
approaches the overdispersed Poisson model) the likelihood may
become multimodal. Therefore, to avoid numerical complications in
actual calculations, we restrict to $p\geqslant1.1$. At the other
extreme, when $p=2$ the number of terms required to evaluate
$c(y;\phi,p)$ may become very large, hence to manage the computation
burden, we restrict $p\leqslant 1.95$. These limitations are also
discussed in Dunn and Smyth (2005). For our data set, we checked
that this restriction did not have a material impact on the results.

\subsection{Maximum likelihood estimation}

The maximum likelihood estimator (MLE) for the parameters is given
by maximizing $L_{\mathcal{D}_{I}}(\bm{\theta})$ in
$\bm{\theta}=(p,\phi,\bm{\alpha },\bm{\beta})$ under the constraints
$\alpha_{i}>0$, $\beta_{j}>0$, $\phi>0$ and $p\in(1,2)$. This leads
to the MLEs
$\widehat{\bm{\theta}}^{\mathrm{MLE}}=(\widehat{p}^{\mathrm{MLE}},
\widehat{\phi}^{\mathrm{MLE}},
\bm{\widehat{{\alpha}}}^{\mathrm{MLE}},
\bm{\widehat{{\beta}}}^{\mathrm{MLE}})$ and to the best estimate
reserves for $R$, given $\mathcal{D}_{I}$,
\begin{equation}
\widehat{R}^{\mathrm{MLE}}=\sum_{i+j>I}\widehat{\alpha}_{i}^{\mathrm{MLE}
}~\widehat{\beta}_{j}^{\mathrm{MLE}}. \label{MLE_ER}
\end{equation}

A convenient practical approach to obtain the MLEs is
to use the fact that at the maximum of the likelihood, $\bm{\beta}$ are
expressed through $\bm{\alpha}$ and $p$ according to the following set of
equations, $p\in\left(  1,2\right)  $:%
\begin{equation}
\beta_{k}=\frac{\sum\limits_{i=0}^{I-k}Y_{i,k}\alpha_{i}^{1-p}}{\sum
\limits_{i=0}^{I-k}\alpha_{i}^{2-p}},\text{ \ \ }k=0,\ldots,I,
\label{betaML2}
\end{equation}
obtained by setting partial derivatives%

\begin{align}
\frac{\partial\ln
L_{\mathcal{D}_{I}}(\bm{\theta})}{\partial\beta_{k}}  &
=\frac{\partial}{\partial\beta_{k}}\sum\limits_{j=0}^{I}\sum\limits_{i=0}
^{I-j}\phi^{-1}\left(
Y_{i,j}\frac{(\alpha_{i}\beta_{j})^{1-p}}{1-p}
-\frac{(\alpha_{i}\beta_{j})^{2-p}}{2-p}\right) \nonumber\\
&  =\sum\limits_{i=0}^{I-k}\phi^{-1}\left(
Y_{i,k}\alpha_{i}^{1-p}\beta
_{k}^{-p}-\alpha_{i}^{2-p}\beta_{k}^{1-p}\right)
\label{LogLHwrtbeta}
\end{align}
equal to zero. Hence, after maximizing the likelihood in $\bm{\alpha
},p,\phi$ one then calculates the set of equations (\ref{betaML2})
for the remaining parameters utilizing the normalization condition
$\alpha_{0}=1$.

Under an asymptotic Gaussian approximation, the distribution of the
MLEs is Gaussian with the covariance matrix elements
\begin{equation}
\text{cov}\left(  \widehat{\theta}_{i}^{\text{MLE}},\widehat{\theta}%
_{j}^{\text{MLE}}\right)  \approx
\left(  \bm{I}^{-1}\right) _{i,j}, \label{MLECorr}%
\end{equation}
where $\bm{I}$ is Fisher's information matrix that can be estimated
by the observed information matrix
\begin{equation}
\left(  \bm{I}\right)  _{i,j} \approx -\left. \frac{\partial^{2}\ln
L_{\mathcal{D}_{I}}\left( \bm{\theta}\right)
}{\partial\theta_{i}\partial\theta_{j}}\right\vert
_{\bm{\theta}{=\widehat{\bm{\theta}}}^{\text{MLE}}}.\label{ObservedInformationMatrix}
\end{equation}

It is interesting to note that,
$\widehat{\beta}_I^{\text{MLE}}=Y_{0,I}$. Also, it is easy to show
(using (\ref{LogLHwrtbeta}) and (\ref{MLECorr})) that
$\widehat{\beta}_I^{\text{MLE}}$ is orthogonal to all other
parameters, i.e.
\begin{equation}
\text{cov}(\widehat{\beta}_{I}^{\text{MLE}},\widehat{\theta}_i^{\text{MLE}})=0,
\text{ \ \ } \widehat{\theta}_i^{\text{MLE}} \ne
\widehat{\beta}_I^{\text{MLE}} \label{ZeroBetaICorr}.
\end{equation}

The next step is to estimate the parameter estimation error in the
reserve as a function of the parameter uncertainty. We do this via
propagation of error by forming a Taylor expansion around the
MLEs, see England and Verrall (2002) formulae (7.6)-(7.8) and
W\"{u}thrich (2003) formulae (5.1)-(5.2),
\begin{align}
\text{stdev}\left(  \widehat{R}^{\mathrm{MLE}}\right)   &  =\sqrt
{\text{Var}\left(  \widehat{R}^{\mathrm{MLE}}\right)  }\label{MLEstderr}\\
\widehat{\text{Var}}\left(  \widehat{R}^{\mathrm{MLE}}\right)   &  =
{\textstyle\sum\limits_{i_{1}+j_{1}>I}}
{\textstyle\sum\limits_{i_{2}+j_{2}>I}}
\widehat{\alpha}_{_{i_{1}}}^{\mathrm{MLE}}\widehat{\alpha}_{_{i_{2}}
}^{\mathrm{MLE}}\text{cov}\left(
\widehat{\beta}_{_{j_{1}}}^{\mathrm{MLE}
},\widehat{\beta}_{_{j_{2}}}^{\mathrm{MLE}}\right) \label{MLE_EE}\\
&  + {\textstyle\sum\limits_{i_{1}+j_{1}>I}}
{\textstyle\sum\limits_{i_{2}+j_{2}>I}}
\widehat{\beta}_{_{j_{1}}}^{\mathrm{MLE}}\widehat{\beta}_{_{j_{2}}
}^{\mathrm{MLE}}\text{cov}\left(
\widehat{\alpha}_{_{i_{1}}}^{\mathrm{MLE}
},\widehat{\alpha}_{_{i_{2}}}^{\mathrm{MLE}}\right) \nonumber\\
&  +2 {\textstyle\sum\limits_{i_{1}+j_{1}>I}}
{\textstyle\sum\limits_{i_{2}+j_{2}>I}}
\widehat{\alpha}_{_{i_{1}}}^{\mathrm{MLE}}\widehat{\beta}_{_{j_{2}}
}^{\mathrm{MLE}}\text{cov}\left(
\widehat{\alpha}_{_{i_{2}}}^{\mathrm{MLE}
},\widehat{\beta}_{_{j_{1}}}^{\mathrm{MLE}}\right)  .\nonumber
\end{align}
Additionally, using the independence assumption on $Y_{i,j}$ and
(2.11), the process variance is estimated as
\begin{align}
\widehat{\text{Var}}\left(  R\right)  =\sum_{i+j>I}\left(
\widehat{\alpha
}_{i}^{\mathrm{MLE}}~\widehat{\beta}_{j}^{\mathrm{MLE}}\right)
^{\widehat {p}^{\mathrm{MLE}}}\widehat{\phi}^{\mathrm{MLE}}.
\label{MLE_PV}
\end{align}

Then the conditional MSEP (\ref{vardecomp}) is estimated by
\begin{align}
\widehat{\mathrm{msep}}_{R|\mathcal{D}_{I}}\left(
\widehat{R}^{\mathrm{MLE}}\right)   & =\widehat{\text{Var}}\left(
R\right)  +\widehat{\text{Var}}\left(
\widehat{R}^{\mathrm{MLE}}\right) \label{decompvarMLE}\\
&  =\text{MLE process variance + MLE estimation error.} \nonumber
\end{align}

Note that, in practice, typically MLE is done for a fixed $p$
(expert choice) and hence model selection questions are neglected.
In our context it means that the expert chooses $p$ and then
estimates $\widehat{\bm{\alpha}}^{\mathrm{MLE}}$,
$\widehat{\bm{\beta}}^{\mathrm{MLE}}$ and
$\widehat{\phi}^{\mathrm{MLE}}$ (see also W\"{u}thrich (2003),
section 4.1). The case $p=1$ corresponds to the overdispersed
Poisson model and provides the chain-ladder estimate for the claims
reserves (see W\"uthrich and Merz (2008), section 2.4). It is
important to note that, often the dispersion parameter $\phi$ is
estimated using Pearson's residuals as
\begin{align}
\widehat{\phi}^{\mathrm{P}}=\frac{1}{N-k}\sum_{i+j\leq I} \frac{
(Y_{i,j}-\widehat{{\alpha}}_{i}^{\mathrm{MLE}}\widehat{{\beta}}_{j}^{\mathrm{MLE}})^2
}{
(\widehat{{\alpha}}_{i}^{\mathrm{MLE}}\widehat{{\beta}}_{j}^{\mathrm{MLE}})^{p}
}, \label{DispersionViaPearson}
\end{align}
where $N$ is the number of observations $Y_{i,j}$ in
${\mathcal{D}_{I}}$ and $k$ is the number of estimated parameters
$\alpha_i$, $\beta_j$ (see e.g. W\"uthrich and Merz (2008), formula
(6.58)). Also note that for a given $p$,
$\widehat{R}^{\mathrm{MLE}}$ given by (\ref{MLE_ER}) does not depend
on $\phi$ and the estimators for the process variance (\ref{MLE_PV})
and estimation error (\ref{MLE_EE}) are proportional to $\phi$. Next
we present the Bayesian model which provides the posterior
distribution of the parameters given the data. This will be used to
analyze the model uncertainty within Tweedie's compound Poisson
models.

\subsection{Bayesian inference}

In a Bayesian context all parameters, $p$, $\phi$, $\alpha_{i}>0$
and $\beta_{j}>0$, are treated as random. Using Bayesian inference
we adjust our \textit{a priori} beliefs about the parameters of the
model utilizing the information from the observations. Through the
Bayesian paradigm we are able to learn more about the distribution
of $p$, $\phi$, $\bm{\alpha}$ and $\bm{\beta}$ after having observed
$\mathcal{D}_{I}$.

Our \textit{a priori }beliefs about the parameters of the model are
encoded in the form of a prior distribution on the parameters
$\pi(\bm{\theta}).$ Then the joint density of
$\mathcal{D}_{I}=\{Y_{i,j}>0;i+j\leq I\}$ and
$\bm{\theta}=\left(p,\phi,\bm{\alpha},\bm{\beta}\right)  $ is given
by
\begin{equation}
\hspace{-1cm}L_{\mathcal{D}_{I}}(\bm{\theta})~\pi(\bm{\theta}).
\end{equation}

Now applying Bayes' law, the posterior distribution of the model parameters,
given the data $\mathcal{D}_{I}$, is%
\begin{equation}
\pi(\bm{\theta}~|~\mathcal{D}_{I})~\propto~L_{\mathcal{D}_{I}}
(\bm{\theta})~\pi(\bm{\theta}). \label{posterior}
\end{equation}
Usually, there are two problems that arise in this context, the
normalizing constant of this posterior is not known in closed form.
Additionally, generating samples from this posterior is typically
not possible using simple inversion or rejection sampling
approaches. In such cases it is usual to adopt techniques such as
Markov chain Monte Carlo (MCMC) methods, see for example Gilks
\textit{et al.} (1996) and Robert and Casella (2004) for detailed
expositions of such approaches.

The Bayesian estimators typically considered are the Maximum a
Postiori (MAP) estimator and the Minimum Mean Square Estimator
(MMSE), that is the mode and mean of the posterior, defined as
follows:
\begin{align}
MAP  &  :\text{ \ \ \
}\hat{\bm{\theta}}^{MAP}=\underset{{\bm{\theta}}}
{\arg\max}\left[  \pi(\bm{\theta}~|~\mathcal{D}_{I})\right]  ,\\
MMSE  &  :\text{ \ \ \ }\hat{\bm{\theta}}^{MMSE}=E\left[
\bm{\theta}~|~\mathcal{D}_{I}\right]  .
\end{align}
We mention here that if the prior $\pi(\bm{\theta})$ is constant and
the parameter range includes the MLE, then the MAP of the posterior
is the same as the MLE. Additionally, one can approximate the
posterior using a second order Taylor series expansion around the
MAP estimate as
\begin{align}
\ln\pi(\bm{\theta}~|~\mathcal{D}_{I})  &
\approx\ln\pi(\hat{\bm{\theta}}
^{MAP}~|~\mathcal{D}_{I})\nonumber\\
&  +\frac{1}{2}\sum_{i,j}\left.
\frac{\partial^{2}}{\partial{\theta}_{i}\partial{\theta}_{j}}
\ln\pi(\bm{\theta}~|~\mathcal{D}_{I})\right\vert _{\bm{\theta}=\hat
{\bm{\theta}}^{MAP}}\left(
{\theta}_{i}-{\hat{\theta}}^{MAP}_{i}\right) \left(
{\theta}_{j}-{\hat{\theta}}^{MAP}_{j}\right).
\label{PosteriorGaussianApprox1}
\end{align}
This corresponds to $\pi(\bm{\theta}~|~\mathcal{D} _{I})$
approximated by the Gaussian distribution with the mean
$\hat{\bm{\theta}}^{MAP}$ and covariance matrix calculated as the
inverse of the matrix
\begin{align}
(\tilde{\bm{I}})_{i,j}=-\left.
\frac{\partial^{2}}{\partial{\theta}_{i}\partial{\theta}_{j}}
\ln\pi(\bm{\theta}~|~\mathcal{D}_{I})\right\vert _{\bm{\theta}=\hat
{\bm{\theta}}^{MAP}}, \label{PosteriorGaussianApprox2}
\end{align}
which in the case of diffuse priors (or constant priors defined on a
large range) compares with the Gaussian approximation for the MLEs
(\ref{MLECorr})-(\ref{ObservedInformationMatrix}).

In the Bayesian context, the conditionally expected future payment,
for Model Assumptions 3.1, is given by
\begin{equation}
E\left.  \left[  R\right\vert \mathcal{D}_{I}\right]
=\sum_{i+j>I}E\left[ \left.  \alpha_{i}\beta_{j}\right\vert
\mathcal{D}_{I}\right]  . \label{gl3}
\end{equation}
Denote the expected reserves, given the parameters $\bm{\theta},$ by
\begin{equation}
\widetilde{R}=E\left[  R|\bm{\theta}\right]  =\sum_{i+j>I}\alpha_{i}\beta_{j}.
\label{rtilda}%
\end{equation}
Then, the best consistent estimate of reserves (ER) is given by
\begin{equation}
\widehat{R}^{\mathrm{B}}=E\left[  \left.  \widetilde{R}~\right\vert
\mathcal{D}_{I}\right]  =\sum_{i+j>I}E\left[  \left. \alpha_{i}\beta
_{j}\right\vert \mathcal{D}_{I}\right]=E\left.  \left[  R\right\vert
\mathcal{D}_{I}\right] ,
\end{equation}
which is, of course, a $\mathcal{D}_{I}$-measurable predictor.
Hence, the conditional MSEP is simply
\begin{equation}
\mathrm{msep}_{R|\mathcal{D}_{I}}\left(  \widehat{R}^{\mathrm{B}}\right)
=E\left[  \left.  \left(  R-\widehat{R}^{\mathrm{B}}\right)  ^{2}\right\vert
\mathcal{D}_{I}\right]  ~=~\mathrm{Var}\left(  \left.  R\right\vert
\mathcal{D}_{I}\right)  . \label{msep}%
\end{equation}
This term, in the Bayesian approach for Tweedie's compound Poisson model, is
decomposed as, see also (\ref{decompvar}),
\begin{align}
\mathrm{Var}\left(  \left.  R\right\vert \mathcal{D}_{I}\right)   &
=\mathrm{Var}\left(  \left.  \sum_{i+j>I}Y_{i,j}\right\vert
\mathcal{D}
_{I}\right)
=\sum_{i+j>I}E\left[  \left.  \left(  \alpha_{i}\beta_{j}\right)
^{p} \phi\right\vert \mathcal{D}_{I}\right]  +\mathrm{Var}\left(
\left.
\widetilde{R}~\right\vert \mathcal{D}_{I}\right)
.
\end{align}
Hence, we obtain the familiar decoupling into
average process variance and estimation
error. However, in addition we incorporate model uncertainty
within Tweedie's compound Poisson model, which enters the
calculation by the averaging over all possible values of the variance
parameter $p$.

\subsection{Random walk Metropolis Hastings-algorithm within Gibbs}

In this section we describe an MCMC method to be used to sample from
the posterior distribution (\ref{posterior}). The following
notations are used: $\bm{\theta}=(p,\phi,\bm{\alpha},\bm{\beta})$ is
the vector of parameters; $U\left(  a,b\right)  $ is the uniform
distribution on the interval $\left(  a,b\right)  $; $f_{N}\left(
x;\mu,\sigma\right)  $ and $F_{N}\left(  x;\mu,\sigma\right)  $ are
the Gaussian density and distribution correspondingly with the mean
$\mu \in \R$ and standard deviation $\sigma>0$ at position $x\in
\R$.

\bigskip

\textbf{Prior Structure:} We assume that all parameters are
independent under the prior distribution $\pi(\bm{\theta})$ and all
distributed uniformly with $\theta_i \sim U\left(  a_i ,b_i\right)$.
 The prior domains we used for our analysis were $p\in\left(
1.1,1.95\right)  $, $\phi\in\left( 0.01,100\right)  $,
$\alpha_{i}\in\left( 0.01,100\right)  $ and $\beta_{j}\in\left(
0.01,10^{4}\right)  $. These are reasonable ranges for the priors in
view of our data in Table \ref{tab2} and corresponding to the
MLEs in Table
\ref{tab3}. Other priors such as diffuse priors can be applied with
no additional difficulty. The choice of very wide prior supports was
made with the aim of performing inference in the setting where the
posterior is largely implied by the data. Subsequently, we checked
that making the ranges wider does not affect the results.

\bigskip

Next we outline a random walk Metropolis-Hastings (RW-MH) within
Gibbs algorithm.\ This creates a reversible Markov chain with the
stationary distribution corresponding to our target posterior
distribution (\ref{posterior}). That is, we will run the chain until
it has sufficiently converged to the stationary distribution
(=posterior distribution) and in doing so we obtain samples from
that posterior distribution. It should be noted that the Gibbs
sampler creates a Markov chain in which each iteration of the chain
involves scanning either deterministically or randomly over the
variables that comprise the target stationary distribution of the
chain. This process involves sampling each proposed parameter update
from the corresponding full conditional posterior distribution. The
algorithm we present generates a Markov chain that will explore the
parameter space of the model in accordance with the posterior mass
in that region of the parameter space. The state of the chain at
iteration $t$ will be denoted by $\bm{\theta}^{t}$ and the chain
will be run for a length of $T$ iterations. The manner in which MCMC
samplers proceed is by proposing to move the $i$th parameter from
state $\theta_i^{t-1}$ to a new proposed state $\theta_i^{\ast}.$
The latter will be sampled from an MCMC proposal transition kernel
(\ref{TransKernel}). Then the proposed move is accepted according to
a rejection rule which is derived from a reversibility condition.
This makes the acceptance probability a function of the transition
kernel and the posterior distribution as shown in
(\ref{acceptProb}). If under the rejection rule one accepts the move
then the new state of the $i$th parameter at iteration $t$ is given
by $\theta_i^t=\theta_i^{\ast}$, otherwise the parameter remains in
the current state $\theta_i^t=\theta_i^{t-1}$ and an attempt to move
that parameter is repeated at the next iteration. In following this
procedure, one builds a set of correlated samples from the target
posterior distribution which have several asymptotic properties. One
of the most useful of these properties is the convergence of ergodic
averages constructed using the Markov chain samples to the averages
obtained under the posterior distribution.

Next we present the algorithm and then some references that will
guide further investigation into this class of simulation
methodology. Properties of this algorithm, including convergence
results can be found in the following references Casella
and George (1992), Robert and Casella (2004), Gelman \textit{et al.}
(1995), Gilks \textit{et al. }(1996) and Smith and Roberts
(1993).

\noindent\hrulefill

\textbf{Random Walk Metropolis Hastings (RW-MH) within Gibbs algorithm.}

1. Initialize randomly or deterministically for $t=0$ the parameter
vector $\bm{\theta}^{0}$ (e.g.~MLEs).

2. For $t=1,\ldots,T$

a) Set $\bm{\theta}^{t}=\bm{\theta}^{t-1}$

b) For $i=1,\ldots,2I+3$

Sample proposal $\theta^{\ast}_i$ from Gaussian distribution whose
density is truncated below $a_i$ and above $b_i$ and given by

\begin{equation}
f_{N}^{T}\left( \theta^{\ast}_i;\theta^{t}_{i},\sigma_{RWi}\right)
=\frac{f_{N}\left(
\theta^{\ast}_i;\theta^{t}_{i},\sigma_{RWi}\right) }{F_{N}\left(
b_i;\theta^{t}_{i},\sigma_{RWi}\right) -F_{N}\left(
a_i;\theta^{t}_{i},\sigma_{RWi}\right)  } \label{TransKernel}
\end{equation}

to obtain
$\bm{\theta}^{\ast}=\left(\theta^{t}_{1},
\ldots,\theta^{t}_{i-1},\theta^{\ast}_i,\theta^{t-1}_{i+1},
\ldots\right)$.

Accept proposal with acceptance probability
\begin{equation}
\alpha\left(\bm{\theta}^{t},\bm{\theta}^{\ast}\right) =\min\left\{
1,\frac{\pi(\bm{\theta}^{\ast}~|~\mathcal{D}_{I})f_{N}^{T}\left(
\theta^{t}_{i};\theta^{\ast}_i,\sigma_{RWi}\right)
}{\pi(\bm{\theta}^{t}
~|~\mathcal{D}_{I})f_{N}^{T}\left(\theta^{\ast}_i;\theta^{t}
_{i},\sigma_{RWi}\right)  }\right\}  , \label{acceptProb}
\end{equation}

where $\pi(\bm{\theta}^{\ast}~|~\mathcal{D}_{I})$ is given by
(\ref{posterior}). That is, simulate $U\sim U(0,1)$ and set
$\theta^t_i=\theta^{\ast}_i$ if $U<\alpha(
\bm{\theta}^t,\bm{\theta}^{\ast}).$

\begin{itemize}
\item[$\Rightarrow$] Note that in (\ref{acceptProb}) the normalizing constant of the
posterior $\pi(\bm{\theta}~|~\mathcal{D}_{I})$ from (\ref{posterior}) is not needed.
\end{itemize}

\noindent\hrulefill

\textbf{Remark.} The RW-MH algorithm is simple in nature and
easily implemented. However, if one does not choose the proposal
distribution carefully, then the algorithm only gives a very slow
convergence to the stationary distribution. There have been
several studies regarding the optimal scaling of proposal
distributions to ensure optimal convergence rates. Gelman
\textit{et al.}~(1997), Bedard and Rosenthal (2007)\ and Roberts
and Rosenthal (2001) were the first authors to publish theoretical
results for the optimal scaling problem in RW-MH algorithms with
Gaussian proposals. For $d$-dimensional target distributions with
i.i.d.~components, the asymptotic acceptance rate optimizing the
efficiency of the process is 0.234 independent of the target
density. In this case we recommend that the selection of
$\sigma_{RWi}$ are chosen to ensure that the acceptance
probability is roughly close to 0.234. This number is the
acceptance probability obtained for asymptotically optimal
acceptance rates for RW-MH algorithms when applied to
multidimensional target distributions with scaling terms possibly
depending on the dimension. To obtain this acceptance rate, one is
required to perform some tuning of the proposal variance prior to
final simulations. An alternative approach is to utilize a new
class of Adaptive MCMC algorithms recently proposed in the
literature, see Atchade and Rosenthal (2005) and Rosenthal (2007),
but these are beyond the scope of this paper.

\subsection{Markov chain results and analysis}

This section presents the results comparing both MLE and Bayesian
estimates for the parameters of Tweedie's compound Poisson model. It
is also demonstrated how additional information in a Bayesian
framework can be obtained through the complete knowledge of the target
posterior distribution obtained from the MCMC algorithm described
above. In this regard we demonstrate how this additional information
can be exploited in the claims reserving setting to provide
alternative statistical analysis not obtainable if one just considers
point estimators. We also analyze model averaging solutions in
section 5. These can be obtained by forming estimates using
the information given by the full posterior distribution $\pi\left(
\bm{\theta}~|~\mathcal{D}_{I}\right)  $ that we find empirically
from the MCMC samples.

The maximum likelihood and MCMC algorithms were implemented in
Fortran. The maximization routine for the MLEs utilizes the direct
search algorithm DBCPOL (that requires function evaluation only)
from the IMSL numerical library. Note that, gradient based optimization
routines such as the BFGS algorithm can be more efficient, but the
direct search algorithm we used was sufficient for our problem in
terms of computing time ($\approx 4$ seconds on a typical desktop
PC\footnotemark[1]).

The algorithm was analyzed on synthetic data and found to provide
correct estimates. In particular with uniform priors the MAP
estimates of the parameters are the same as the MLEs, up to
numerical errors. This was confirmed for different sized claims
triangles. The actual data set studied in this paper is presented
in  Table \ref{tab2}. The data we study is
 the standard data set used in W\"uthrich and Merz (2008) scaled by 10,000.

The results presented for the Bayesian approach were obtained after
pretuning the Markov chain random walk standard deviations,
$\sigma_{RW_{i}},$ to produce average acceptance probabilities of
$0.234.$ Then the final simulation was for $10^5$ iterations from a
Markov chain ($\approx 17$min\footnotemark[1]) in which the first
$10^4$ iterations were discarded as burnin when forming the
estimates. \footnotetext[1]{Intel$^\text{\textregistered}$
Core$^{TM}$2 Duo, 2.13GHz processor.} The pretuned proposal standard
deviations $\sigma_{RW_i}$ are presented in Table \ref{tab3}. The
first set of results in Table \ref{tab3} demonstrates the MLE versus
the Bayesian posterior estimator MMSE for all model parameters.
Included are the [5\%, 95\%] predictive intervals for the Bayesian
posterior distribution. The MLE standard deviations are calculated
using (\ref{MLECorr}). The numerical standard errors (due to a
finite number of MCMC iterations) in the Bayesian estimates are
obtained by blocking the MCMC samples post burnin into blocks of
length 5000 and using the estimates on each block to form the
standard error (these are given in brackets next to the estimates).

\begin{figure}[b]
\centerline{\includegraphics{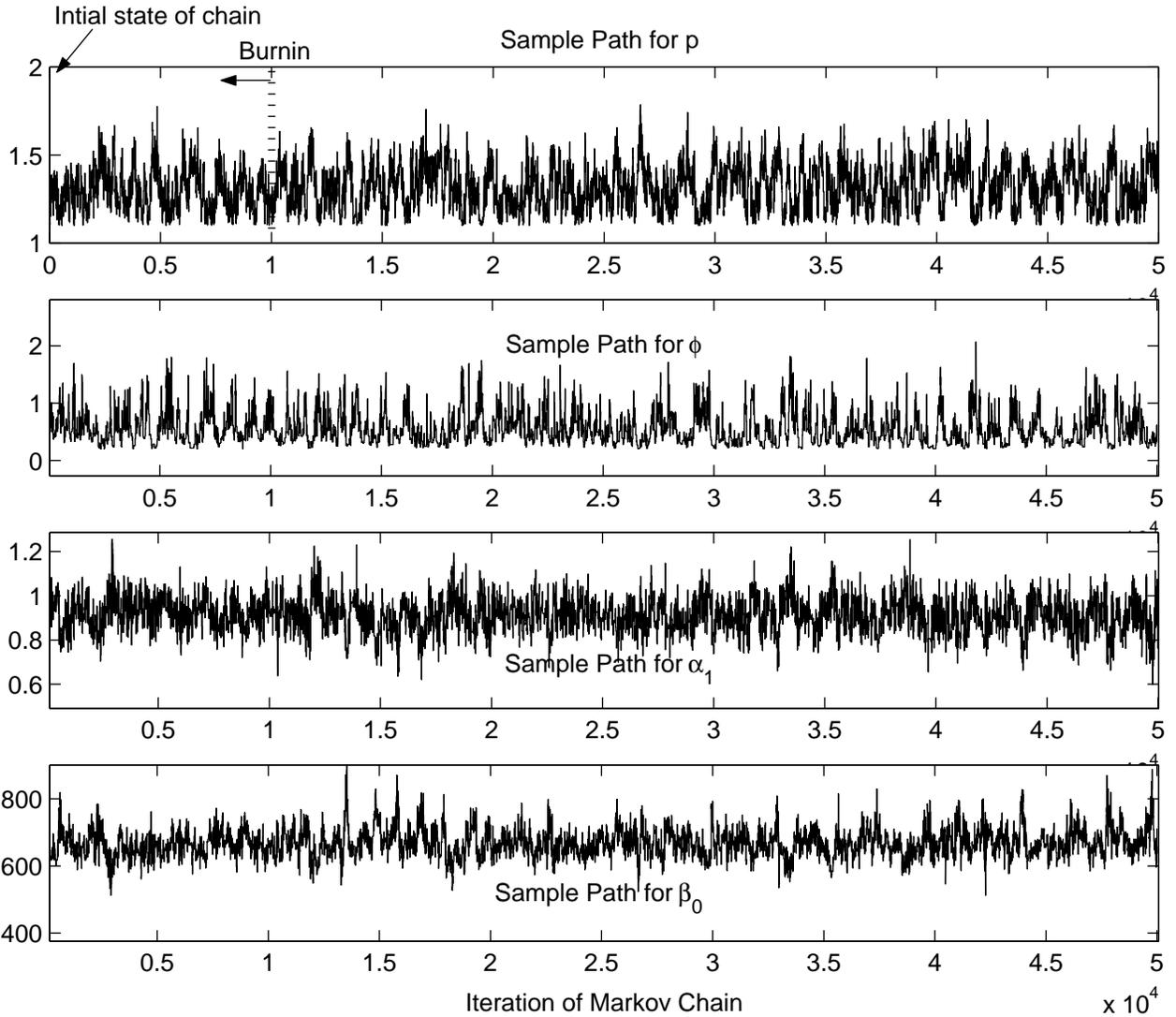}}
\caption{Markov chain sample paths $\left(
p,\phi,\alpha_{1},\beta_{0}\right)  $.}
\label{fig1}
\end{figure}

\begin{figure}[ptb]
\centerline{\includegraphics{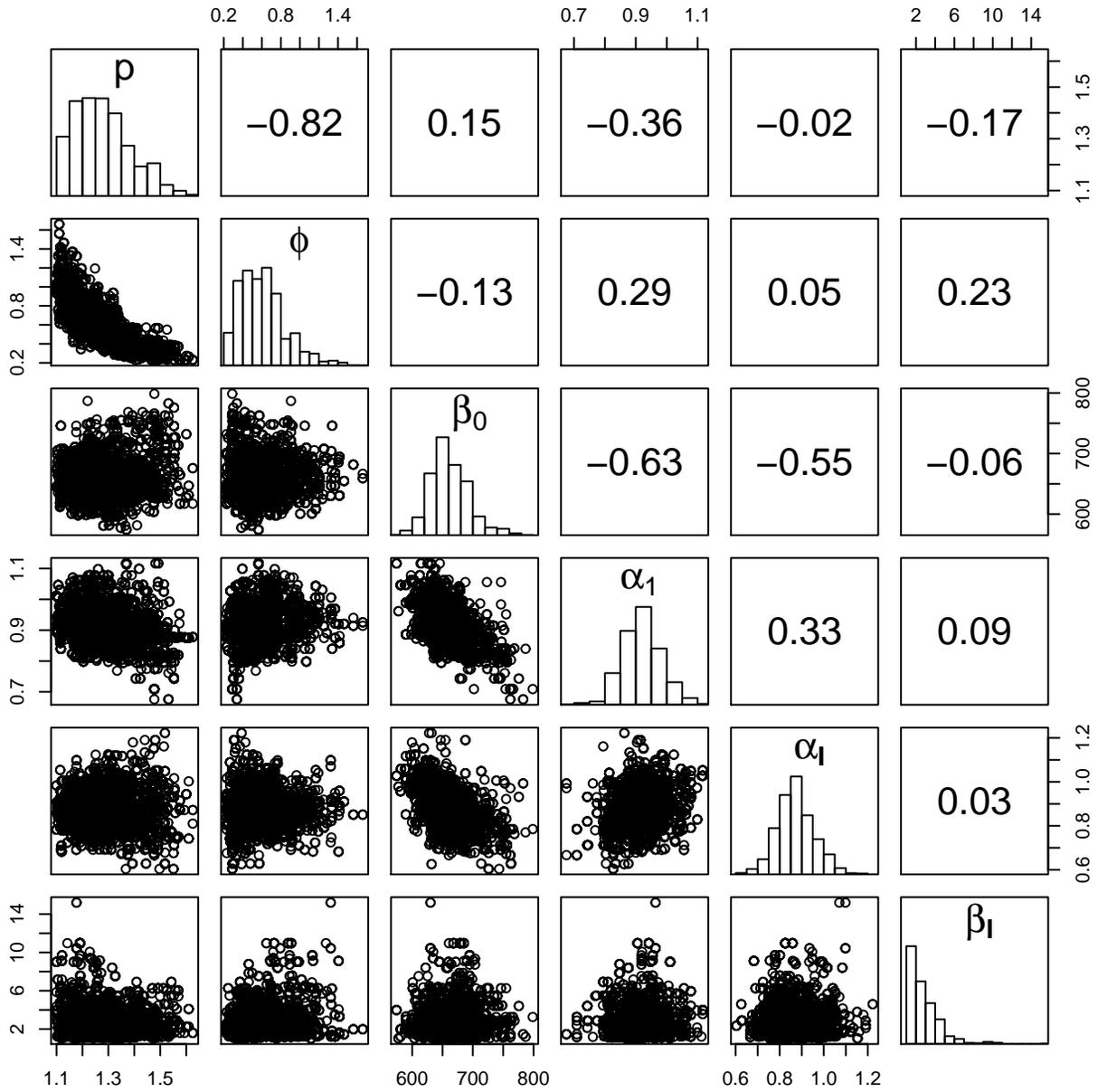}}.\caption{Posterior
scatter plots, marginal posterior histograms and linear correlations
for $\left( p,\phi,\alpha_{1},\beta_{0},\alpha_{I},\beta_{I}\right)
.$}
\label{fig2}
\end{figure}

The next set of analysis demonstrates the performance of the MCMC
approach in converging to the stationary distribution given by the
target posterior $\pi(\bm{\theta}~|~\mathcal{D}_{I})$. To analyze
this, in Figure \ref{fig1}, we present the trace plots for the
Markov chain for the parameters, $\left(
p,\phi,\alpha_{1},\beta_{0}\right)  $. Also, in Figure \ref{fig2},
we demonstrate the marginal posterior distribution histograms and
pair-wise posterior scatter plots for $\left(
p,\phi,\alpha_{1},\beta_{0},\alpha _{I},\beta_{I}\right)$. The
lower panels in Figure \ref{fig2} are the scatter plots for the
pair-wise marginal posteriors, the diagonal contains the marginal
posteriors and the upper panels contains the correlations between
parameters. These plots demonstrate strong linear correlations
between several parameters. Some of these correlations are similar
to MLE correlations calculated using (\ref{MLECorr}). For example,
we found that under the posterior distribution $\rho(p,\phi)
\approx-0.82$ and $\rho(\beta_0,\alpha_1) \approx-0.63$, see
Figure \ref{fig2}, are similar to $\rho(
\widehat{p}^{\text{MLE}},\widehat{\phi}^{\text{MLE} })
\approx-0.94$ and $\rho( \widehat{\beta}_{0}^{\text{MLE}
},\widehat{\alpha}_{1}^{\text{MLE}})  \approx-0.68$
correspondingly. However, we also observed that under the
posterior distribution $\rho( p,\beta_{I}) \approx -0.17 $ and
$\rho(\phi,\beta_{I})\approx 0.23$, see Figure \ref{fig2}, while
corresponding MLE correlations are zero, see
(\ref{ZeroBetaICorr}).

\section{Variable selection via posterior model probabilities}

In the development so far it has been assumed that variable
selection is not being performed, that is we are assuming that the
model is known and we require parameter estimates for this model.
This is equivalent to specifying that the number of $\alpha$ and
$\beta$ parameters is fixed and known in advance. We now relax this
assumption and will demonstrate how the variable selection problem
can be incorporated into our framework. The procedure we utilize for
the variable selection is based on recent work of Congdon (2006) and
specifies the joint support of the posterior distribution for the
models and parameters under the product space formulation of Carlin
and Chib (1995).

In this section we consider the subset of nested models which create
homogenous blocks in the claims reserving triangle $(I=9)$ for the
data set in Table 2.

\begin{itemize}
\item $M_{0}:\bm{\theta}_{[0]}=\left(p,\phi,\widetilde
{{\alpha}}_{0}={\alpha}_{0},\ldots,\widetilde{{\alpha}}_{I}={\alpha}
_{I},\widetilde{{\beta}}_{0}={\beta}_{0},\ldots,\widetilde{{\beta}}_{I}={\beta
}_{I}\right)  $ - \textbf{saturated model}.

\item $M_{1}:\bm{\theta}_{[1]}=\left(p,\phi,\widetilde{{\beta}}_{0}\right)
$ with $\left(
\widetilde{{\beta}}_{0}={\beta}_{0}=\ldots={\beta}_{I}\right) ,\left(
{\alpha}_{0}=\ldots={\alpha}_{I}=1\right)  .$

\item $M_{2}:\bm{\theta}_{[2]}=\left(p,\phi,\widetilde{{\alpha}}_{1},\widetilde
{{\beta}}_{0},\widetilde{{\beta}}_{1}\right)  $ with $\left({\alpha
}_{0}=\ldots={\alpha}_{4}=1\right)  $, $\left(
\widetilde{{\alpha}}_{1}={\alpha }_{5}=\ldots={\alpha}_{I}\right)
$,\newline$\left(  \widetilde{{\beta}}
_{0}={\beta}_{0}=\ldots={\beta}_{4}\right)  $, $\left(
\widetilde{{\beta}} _{1}={\beta}_{5}=\ldots={\beta}_{I}\right)  $.

\item $M_{3}:\bm{\theta}_{[3]}=\left(p,\phi,\widetilde{{\alpha}}_{1},\widetilde
{{\alpha}}_{2},\widetilde{{\beta}}_{0},\widetilde{{\beta}}_{1},\widetilde
{{\beta}}_{2}\right)  $ with $\left(  {\alpha}_{0}={\alpha}
_{1}=1\right)  $, $\left(
\widetilde{{\alpha}}_{1}={\alpha}_{2}=\ldots={\alpha }_{5}\right)
$,\newline$\left(  \widetilde{{\alpha}}_{2}={\alpha}
_{6}=\ldots={\alpha}_{I}\right)  $, $\left(
\widetilde{{\beta}}_{0}={\beta} _{0}={\beta}_{1}\right)  $, $\left(
\widetilde{{\beta}}_{1}={\beta} _{2}=\ldots={\beta}_{5}\right)  $,
$\left( \widetilde{{\beta}}_{2}={\beta} _{6}=\ldots={\beta}_{I}\right)
.$

\item $M_{4}:\bm{\theta}_{[4]}=\left(p,\phi,\widetilde{{\alpha}}_{1},\widetilde
{{\alpha}}_{2},\widetilde{{\alpha}}_{3},\widetilde{{\beta}}_{0},\widetilde
{{\beta}}_{1},\widetilde{{\beta}}_{2},\widetilde{{\beta}}_{3}\right)
$ with $\left(  {\alpha}_{0}={\alpha}_{1}=1\right)  $, $\left(
\widetilde {{\alpha}}_{1}={\alpha}_{2}={\alpha}_{3}\right)  $,

$\left(  \widetilde{{\alpha}}_{2}={\alpha}_{4}={\alpha}_{5}={\alpha}
_{6}\right)  $, $\left(
\widetilde{{\alpha}}_{3}={\alpha}_{7}={\alpha}
_{8}={\alpha}_{I}\right)  $, $\left( \widetilde{{\beta}}_{0}={\beta}
_{0}={\beta}_{1}\right)  $, $\left(  \widetilde{{\beta}}_{1}={\beta}
_{2}={\beta}_{3}\right)  $, \newline$\left(
\widetilde{{\beta}}_{2}={\beta }_{4}={\beta}_{5}={\beta}_{6}\right)
$, $\left(  \widetilde{{\beta}}
_{3}={\beta}_{7}={\beta}_{8}={\beta}_{I}\right)  .$

\item $M_{5}:\bm{\theta}_{[5]}=\left(p,\phi,\widetilde{{\alpha}}_{1},\widetilde
{{\alpha}}_{2},\widetilde{{\alpha}}_{3},\widetilde{{\alpha}}_{4}
,\widetilde{{\beta}}_{0},\widetilde{{\beta}}_{1},\widetilde{{\beta}}
_{2},\widetilde{{\beta}}_{3},\widetilde{{\beta}}_{4}\right) $ with
$\left(  {\alpha}_{0}={\alpha}_{1}=1\right)  $, $\left(
\widetilde{{\alpha} }_{1}={\alpha}_{2}={\alpha}_{3}\right)  $,
$\left(  \widetilde{{\alpha}} _{2}={\alpha}_{4}={\alpha}_{5}\right)
$, $\left(  \widetilde{{\alpha}}
_{3}={\alpha}_{6}={\alpha}_{7}\right)  $, $\left(
\widetilde{{\alpha}} _{4}={\alpha}_{8}={\alpha}_{I}\right)  $,
$\left(  \widetilde{{\beta}} _{0}={\beta}_{0}={\beta}_{1}\right)  $,
$\left(  \widetilde{{\beta}} _{1}={\beta}_{2}={\beta}_{3}\right)  $,
$\left(  \widetilde{{\beta}} _{2}={\beta}_{4}={\beta}_{5}\right)  $,
$\left(  \widetilde{{\beta}} _{3}={\beta}_{6}={\beta}_{7}\right)  $,
$\left(  \widetilde{{\beta}} _{4}={\beta}_{8}={\beta}_{I}\right)  .$

\item $M_{6}:\bm{\theta}_{[6]}=\left(p,\phi,{\alpha}_{0},\widetilde{{\alpha}}
_{1},{\beta}_{0},{\beta}_{1},\ldots,{\beta}_{I}\right)  $ with $\left(
\widetilde{{\alpha}}_{1}={\alpha}_{1}=\ldots={\alpha}_{I}\right)  .$
\end{itemize}

Now, to determine the optimal model, we first consider the joint
posterior distribution for the model probability and the model
parameters denoted $\pi(M_{k},\bm{\theta}_{[k]}~|~\mathcal{D}_{I}),$
where $\bm{\theta}_{[k]} =\left(
\widetilde{{\theta}}_{1,[k]},\widetilde{{\theta}}_{2,[k]},\ldots,\widetilde
{{\theta}}_{N_{\left[k\right]},[k]}\right)  $ is the parameter
vector for model $[k].$ Additionally we denote the prior bounds for
$\widetilde{{\theta} }_{i,[k]}$ as $\left[
a_{\widetilde{{\theta}}_{i,[k]}},b_{\widetilde{{\theta
}}_{i,[k]}}\right]  .$ We assume a prior distribution $\pi\left(
M_{k}\right)  $ for the model selection and a prior for the
parameters conditional on the model $\pi\left(
\bm{\theta}_{[k]}~|~M_{k}\right) $. It is no longer possible to run
the standard MCMC procedure we described in section 3.4 for this
variable selection setting. This is because the posterior is now
defined on either a support consisting of disjoint unions of
subspaces or a product space of all such subspaces, one for each
model considered. A popular approach to run Markov chains in such a
situation is to develop a more advanced sampler than that presented
above, typically in the disjoint union setting. This involves
developing a Reversible Jump RJ-MCMC framework, see Green (1995) and
the references therein. This type of Markov chain sampler is
complicated to develop and analyze. Hence, we propose as an
alternative in this paper to utilize a recent procedure that will
allow us to use the above MCMC sampler we have already developed for
a model $M_{k}.$ The process we must follow involves first running
the sampler in the simulation technique described in section 3.4 for
each model considered. Then the calculation of the posterior model
probabilities $\pi(M_{k}~|~\mathcal{D}_{I})$ is performed using the
samples from the Markov chain in each model to estimate
(\ref{ModProbsPost}).

Furthermore, our approach here removes the assumption on the priors
across models, made by Congdon (2006), p.348,
\begin{equation}
\pi\left(  \bm{\theta}_{\left[  m\right]  }~|~M_{k}\right)  =1,m\neq k
\end{equation}
and instead we work with the prior
\begin{equation}
\pi(\bm{\theta}_{\left[  m\right]
}~|~M_{k})={\textstyle\prod\limits_{i=1} ^{N_{\left[  m\right]
}}}\left[  b_{\widetilde{{\theta}}_{i,[m]}
}-a_{\widetilde{{\theta}}_{i,[m]}}\right]  ^{-1},m\neq k.
\end{equation}

That is, instead we use a class of priors where specification of
priors for a model $M_{k}$ automatically specifies priors for any
other model. This is a sensible set of priors to consider given our
product space formulation and it has a clear interpretation in our
setting where we specify our models through a series of constraints,
relative to each other. In doing this we also achieve our goal of
having posterior model selection insensitive to the choice of the
prior and being data driven. The modified version of Congdon's
(2006), formula A.3, we obtain after relaxing Congdon's assumption,
allows the calculation of the posterior model probabilities
$\pi(M_{k}~|~\mathcal{D}_{I})$ using the samples from the
Markov chain in each model to estimate%
\begin{align}
\pi(M_{k}~|~\mathcal{D}_{I})  &  =\int\pi(M_{k},\bm{\theta}_{[k]}
~|~\mathcal{D}_{I})d\bm{\theta}_{[k]}
=\int\pi(M_{k}~|~\bm{\theta}_{[k]},\mathcal{D}_{I})\pi(\bm{\theta}_{[k]}
~|~\mathcal{D}_{I})d\bm{\theta}_{[k]}\nonumber\\
&
\approx\frac{1}{T-T_{b}}\sum\limits_{j=T_{b}+1}^{T}\pi(M_{k}~|~\mathcal{D}
_{I},\bm{\theta}_{j,[k]})\nonumber\\
&
=\frac{1}{T-T_{b}}\sum\limits_{j=T_{b}+1}^{T}\frac{L_{\mathcal{D}_{I}
}(M_{k},\bm{\theta}_{j,[k]}) {\textstyle\prod\limits_{k=0}^{K}}
\pi(\bm{\theta}_{j,[k]}~|~M_{k})\pi(M_{k})}{\sum\nolimits_{m=0}^{K}
L_{\mathcal{D}_{I}}(M_{m},\bm{\theta }_{j,[m]})
{\textstyle\prod\limits_{k=0}^{K}}
\pi(\bm{\theta}_{j,[k]}~|~M_{m})\pi(M_{m})}\nonumber\\
&
=\frac{1}{T-T_{b}}\sum\limits_{j=T_{b}+1}^{T}\frac{L_{\mathcal{D}_{I}
}(M_{k},\bm{\theta}_{j,[k]})}{\sum\nolimits_{m=0}^{K}L_{\mathcal{D}_{I}}
(M_{m},\bm{\theta }_{j,[m]})}. \label{ModProbsPost}
\end{align}
Here $K=6,$ and for a proof, see Congdon (2006), formula A.3. Note
that, the prior of parameters (given model) contributes in the
above implicitly as $\bm{\theta}_{j,[k]}$ are MCMC samples from
the $k^{th}$ models posterior distribution. In the actual
implementation we used $T=100,000$ and the burnin period
$T_{b}=10,000.$ Note, the prior probabilities for each model are
considered diffuse and are set such that all models \textit{a
priori }are equiprobable, hence $\pi(M_{k})=1/\left(  K+1\right) $
and $\pi(\bm{\theta}_{j,[k]}~|~M_{k})$ is the prior for model
$M_{k}$'s parameters evaluated at the $j^{th}$ Markov chain
iteration. Once we have the posterior model probabilities we can
then take the MAP estimate for the optimal model (variable
selection) for the given data set. In this paper we do not
consider the notion of model averaging over different
parameterized models in the variable selection context. Instead we
simply utilize these results for optimal variable selection from a
MAP perspective for the marginal posterior
$\pi(M_{k}~|~\mathcal{D}_{I})$.

In addition to this model selection criterion we also consider in
the Bayesian framework the Deviance Information Criterion (DIC), see
Bernardo and Smith (1994). From a classical maximum likelihood
perspective we present the likelihood ratio (LHR) p-values.

Application of this technique to the simulated MCMC samples for each
of the considered models produced the posterior model probabilities
given in Table \ref{tab4}. This suggests that within this subset of
models considered, the saturated model $M_{0}$ was the optimal model
to utilize in the analysis of the claims reserving problem,
$\pi\left(  M_{0}~|~\mathcal{D}_{I}\right) \approx0.7$. It is
followed by model $M_6$ with $\pi\left(
M_{0}~|~\mathcal{D}_{I}\right) \approx 0.3$. Additionally, the
choice of $M_{0}$ was also supported by the other criteria we
considered: DIC and LHR.

In future research it would be interesting to extend to the full
model space which considers all models in the power set $\left\vert
\bm{\theta}_{[0]} \right\vert $. This is a large set of models
including all combinatorial combinations of model parameters for
$\alpha^{\prime}s$ and $\beta^{\prime}s$. In such cases it is no
longer feasible to run standard MCMC algorithms in each model since this will
involve an impractical number of simulations. Hence, more
sophisticated model exploration techniques will be required such as
RJ-MCMC, see Green (1995) or the product space samplers of Carlin and
Chib (1995).

We note here that we do not claim $M_{0}$ is the optimal model in all possible
models, only in the subset we consider in this section. In saying this we
acknowledge that we aim to work in the saturated model but consider it
important to illustrate how variable selection can be performed in this class
of models and also raise awareness that this will impact the model
uncertainty
analysis subsequently performed.

Hence, using these findings and the analysis of the MCMC results for model
$M_{0}$ provided above, we may now proceed to analyze the claims reserving
problem. Of \ interest to the aim of this paper is the sensitivity of the
model choice parameter $p$ to the parameterization of the claims reserving
triangle. This is particularly evident when one considers the MMSE estimate of
the model specification parameter $p$ estimated under each model. In the most
parsimonious, yet inflexible model $M_{1}$ the estimate obtained was
$MMSE\left(  p\right)  \approx1.9$, a very similar estimate was obtained in
models $M_{2},M_{3},M_{4}$ and $M_{5},$ however, interestingly in the
saturated model the estimate was $MMSE\left(  p\right)  \approx1.3$ which is
almost at the other extreme of the considered range for which the parameter
$p$ is defined.

\section{Calculation of the claims reserves}

We now demonstrate the results for several quantities in the claims
reserving setting, utilizing the MCMC simulation results we obtained
for the Bayesian posterior distribution under the variable selection
model $M_0$ (saturated model). In particular, we start by noting
that we use uniform prior distributions with a very wide ranges to
perform inference implied by the data only. In this case,
theoretically, the Bayesian MAP (the posterior mode) and MLEs for
the parameters should be identical up to numerical error due to the
finite number of MCMC iterations. A large number of MCMC iterations
was performed so that the numerical error is not material. In
general, the use of more informative priors will lead to the
differences between the MAP and MLE. Some of the MMSE estimates (the
posterior mean) were close to the MAP estimates, indicating that the
marginal posterior distributions are close to symmetric. When the
posterior is not symmetric, MMSE and MAP can be very different.
Also, note that the uncertainties in the parameter MLEs are
estimated using the asymptotic Gaussian approximation
(\ref{MLECorr})-(\ref{ObservedInformationMatrix}). In the case of
constant priors, this should lead to the same inferences as
corresponding Bayesian estimators if the posterior distributions are
close to the Gaussian approximation, see
(\ref{PosteriorGaussianApprox1})-(\ref{PosteriorGaussianApprox2}).
In addition, the MLEs for the reserves, estimation error and process
variance, see section 3.2, are based on a Taylor expansion around
parameter MLEs assuming small errors. In many cases the posterior is
materially different from the Gaussian distribution, has significant
skewness and large standard deviation leading to the differences
between the  MLEs and corresponding Bayesian estimators. Having
mentioned this, we now focus on the main point of this paper which
involves analysis of the quantities in Table \ref{tab5} related to
the model uncertainty within Tweedie's compound Poisson models
(introduced by fixing model parameter $p$) in a Bayesian setting.

It is worth noting that  point estimates of model parameters are
either in the frequentists approach MLEs or in a Bayesian approach
the MAP or MMSE estimates. These are under the auspice that we wish
to perform model selection (i.e.~selection of $p$). The focus of
this paper is to demonstrate the difference in results obtained for
reserve estimates that can arise by performing model averaging
instead of the typical approach of model selection, using \textit{a
priori} chosen $p$. In this regard we perform estimation utilizing
the full posterior distribution of the parameters and not just point
estimators. This allows us to capture the influence of the model
uncertainty (uncertainty in $p$), since in a Bayesian setting we can
account for this uncertainty using the posterior distribution. In
particular, the Bayesian analysis specifies the optimal $p$ (either
in the MAP or the MMSE context) and it also provides a confidence
interval for the choice of $p$ (see Figure \ref{fig7}), which
corresponds to the choice of the optimal model within Tweedie's
compound Poisson models. Moreover, we demonstrate the impact on the
claims reserve by varying $p$ from 1.1 to 1.9 (i.e.~for a fixed
model choice).

\subsection{Results: average over $p$}

\begin{figure}[ptb]
\centerline{\includegraphics{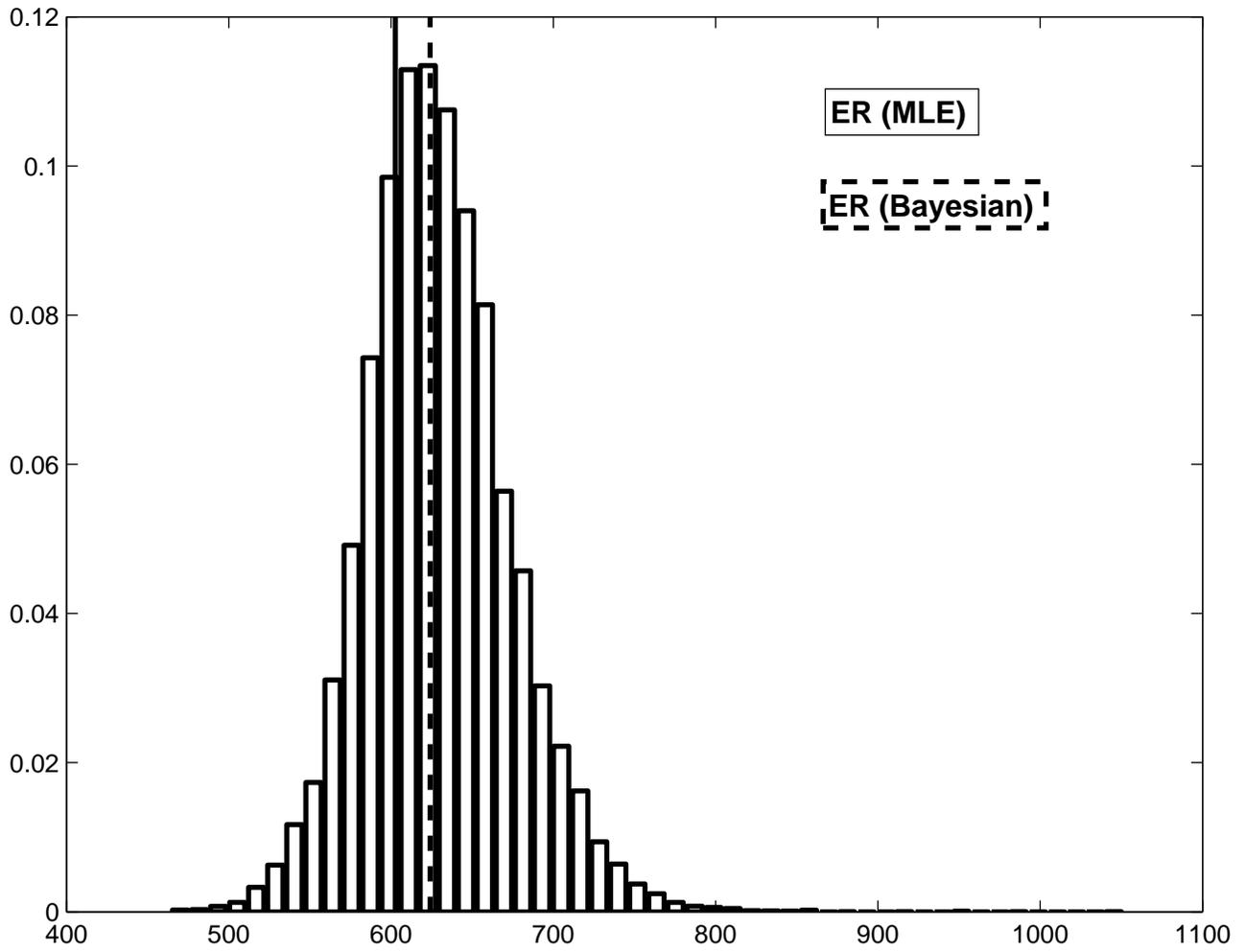}}\caption{Predicted
distribution of reserves,
$\widetilde{R}=\sum\limits_{i+j>I}\alpha_{i} \beta_{j}.$}
\label{fig3}
\end{figure}

\begin{figure}[ptb]
\centerline{\includegraphics{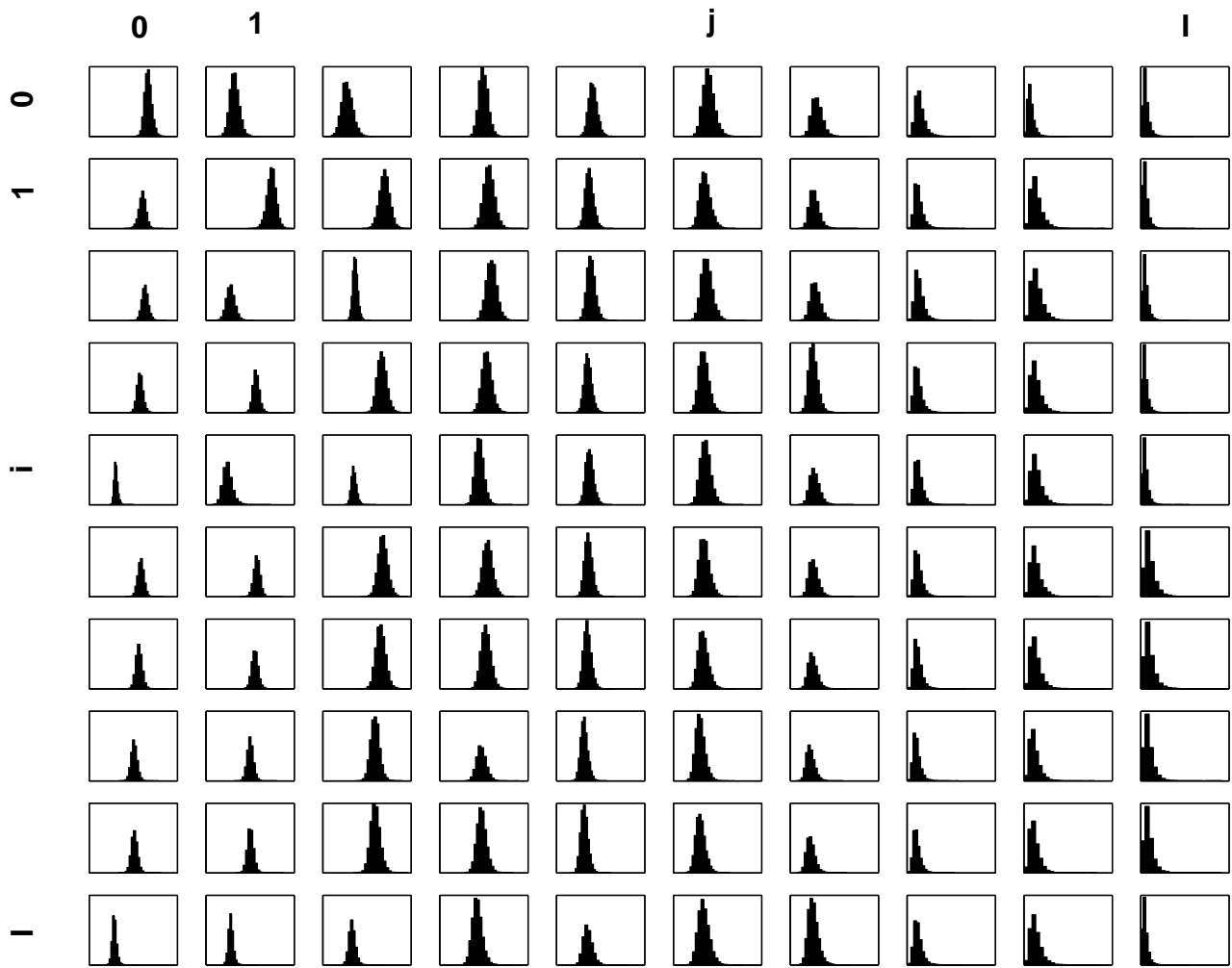}}\caption{Posterior
distributions for $\widetilde{R}_{i,j}=\alpha_{i} \beta_{j}$
estimated using MCMC.}
\label{fig4}
\end{figure}

Initially it is worth considering the predicted reserve distribution
for the estimator $\widetilde{R}$. This is obtained by taking the
samples $t=10,001$ to $100,000$ from the MCMC simulation $\left\{
p^{t},\phi^{t},\bm{\alpha}^{t} ,\bm{\beta}^{t}\right\}  $ and
calculating $\left\{  \widetilde{R}^{t}\right\}  $ via
(\ref{rtilda}). The histogram estimate is presented in Figure
\ref{fig3}. In the same manner, we also estimate the distributions
of $\widetilde{R}_{i,j}=\alpha_{i}\beta_{j}$ for the individual
cells of the $I\times I$
claims matrix, presented as subplots in Figure \ref{fig4}. Note that
the total observed loss in the upper triangle ($\approx 9274$) is
consistent with $E[\sum\limits_{i+j\leq I}\alpha_{i}\beta_{j}]$ and
$[{\text{Var}(\sum\limits_{i+j\leq I}\alpha_{i}\beta_{j})}]^{1/2}$
estimated using the MCMC samples as ($\approx9311$) and
($\approx190$) respectively. The maximum likelihood approach results
in $\sum\limits_{i+j\leq I}
\hat{\alpha}_{i}^{MLE}\hat{\beta}_{j}^{MLE}\approx 9275$ with
standard deviation $\approx 124$ also conforming with the observed
total loss.

Now we focus on quantities associated with the estimated
distribution for $\widetilde{R}~$ to calculate the results, see
Table \ref{tab5}, which can only be estimated once the entire
posterior distribution is considered. These quantities are the key
focus of this paper since they allow assessment of the conditional
MSEP as specified in (\ref{msep}). In particular, we may now
easily use the posterior probability samples obtained from the
MCMC algorithm to evaluate the estimated reserve (ER), the process
variance (PV) and the estimation error (EE) in the conditional
MSEP. This provides an understanding and analysis of the behaviour
of the proposed model in both the model averaging and model
selection (i.e.~selection of $p$) contexts whilst considering the
issue of model uncertainty, the goal of this paper. The Bayesian
estimates for ER, PV, EE and MSEP are presented in Table
\ref{tab6}. The corresponding MLEs were calculated using
(\ref{MLE_ER}), (\ref{MLE_PV}), (\ref{MLE_EE}) and
(\ref{decompvarMLE}) respectively and presented in Table
\ref{tab6} for comparison. The results demonstrate the following:

\begin{itemize}
\item Claims reserves MLE, $\widehat{R}^{\mathrm{MLE}},$ is less than Bayesian estimate
$\widehat{R}^{\mathrm{B}}$ by approximately 3\%, which is the
estimation bias of the claims reserve MLE (see also
W\"uthrich and Merz (2008), Remarks 6.15.

\item $\sqrt{EE}$ and $\sqrt{PV}$ are of the same magnitude,
approximately 6-7\% of the total claims reserves.

\item MLEs for $\sqrt{EE}$ and $\sqrt{PV}$ are less than corresponding Bayesian
estimates by approximately 37\% and 30\%, respectively.

\item The difference between $\widehat{R}^{\mathrm{MLE}}$ and $\widehat
{R}^{\mathrm{B}}$ is of the same order of magnitude as $\sqrt{EE}$
and $\sqrt{PV}$ and thus is significant.
\end{itemize}

Note that we use constant priors with very wide ranges, the MLE
uncertainties are calculated using an asymptotic Gaussian
approximation and numerical error due to the finite number of MCMC
iterations is not material (also see the 1st paragraph, section 5).
The observed significant differences between the MLEs and
corresponding Bayesian estimators suggest that our posterior
distributions are skewed and materially different from the Gaussian
distribution.

\begin{figure}[ptb]
\centerline{\includegraphics{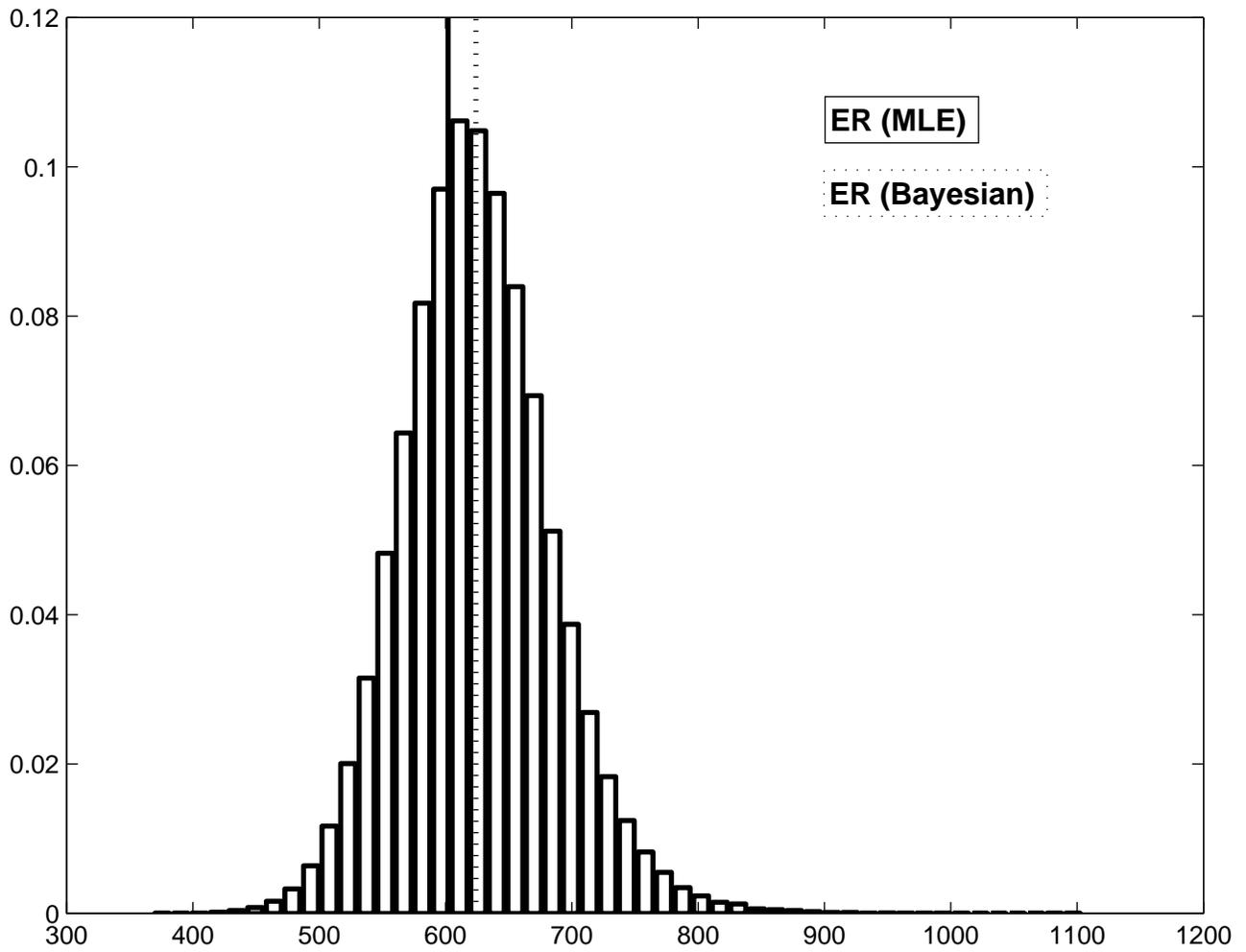}}\caption{
Distribution of total outstanding claims payment
$R={\textstyle\sum\limits_{i+j>I}}Y_{i,j}$, accounting for all
process, estimation and model uncertainties. }
\label{fig5}
\end{figure}

We conclude this section with the distribution of $R$, the total
outstanding claims payment, see Figure \ref{fig5}. This is
obtained from the MCMC samples of the parameters
$(p,\phi,\bm{\alpha },\bm{\beta})$ which we then transform to
parameters $\left( \bm{\lambda},\gamma ,\bm{\tau}\right)$ from
model representation 1, section 2, and simulate annual losses in
$i+j>I$. That is, these samples of $R$ are obtained from the full
predictive distribution $f\left( \left. R~\right\vert
\mathcal{D}_{I}\right) = {\textstyle\int} g\left( \left.
R~\right\vert \bm{\theta}\right) \pi\left(  \left.
\bm{\theta}~\right\vert \mathcal{D}_{I}\right) d\bm{\theta},$
where $g\left( \left. R~\right\vert \bm{\theta}\right)  $ is the
distribution of $R$ given by (\ref{R}) and (\ref{Tweedie Model}).
It takes into account both process uncertainty and parameter
uncertainty. We note that while reserving by some measure of
centrality such as $\widehat{R}^{\mathrm{B}}$ may be robust, it
will not take into account the distributional shape of $R$. A
viable alternative may be Value-at-Risk (VaR) or a coherent risk
measure such as Expected Shortfall. In Table \ref{tab7} we
demonstrate estimates of the VaR for $\widetilde{R}$ and $R$ at the $
75\%,90\%$ and $95\%$ quantiles.

\subsection{Results: conditioning on $p$}

\begin{figure}[ptb]
\centerline{\includegraphics{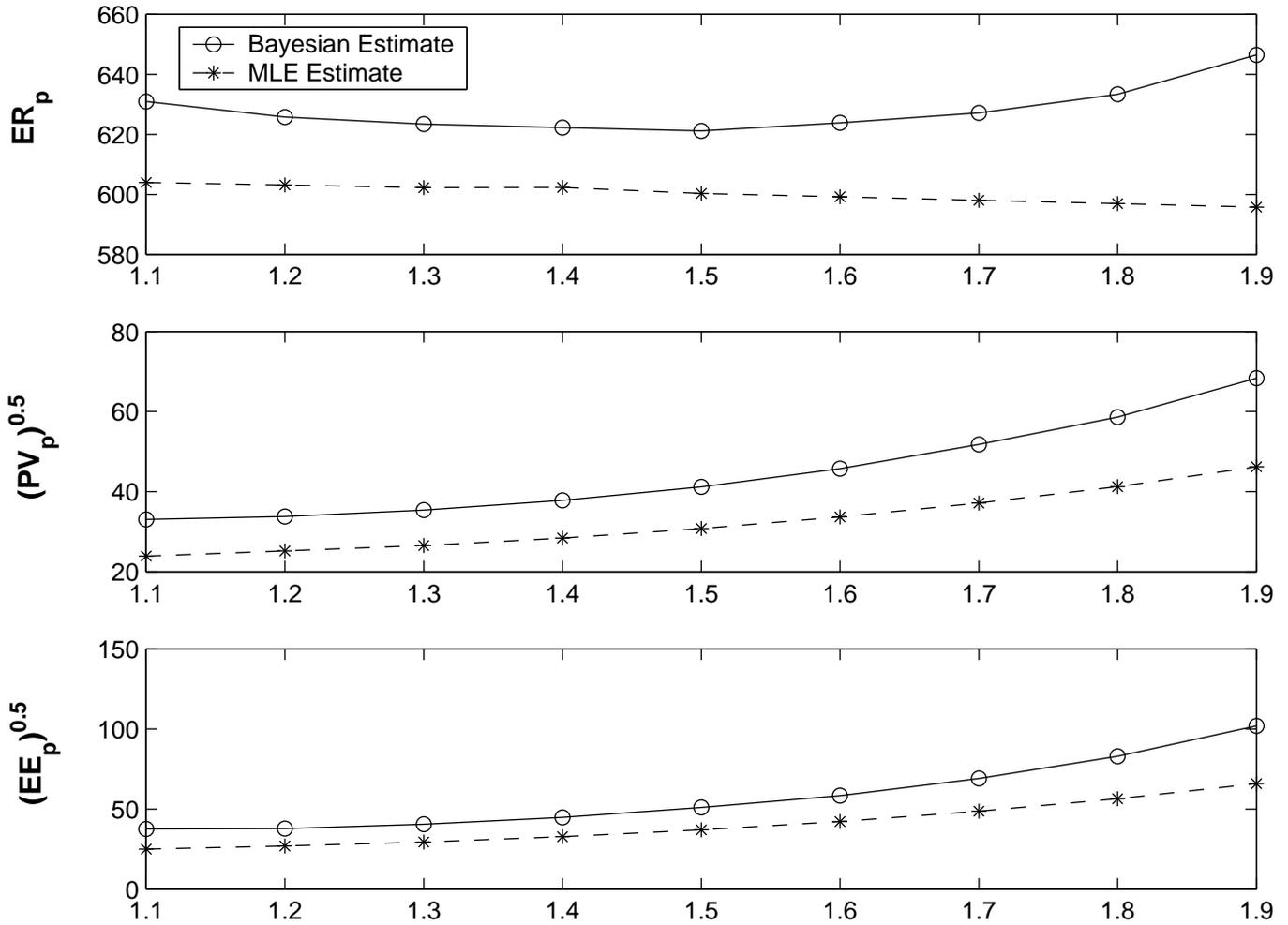}}\caption{Estimates
of quantities from Table \ref{tab5} conditional on $p$. Note,
numerical standard errors are not included as they are negligible
and are less than the size of the symbols. }
\label{fig6}
\end{figure}

\begin{figure}[ptb]
\centerline{\includegraphics{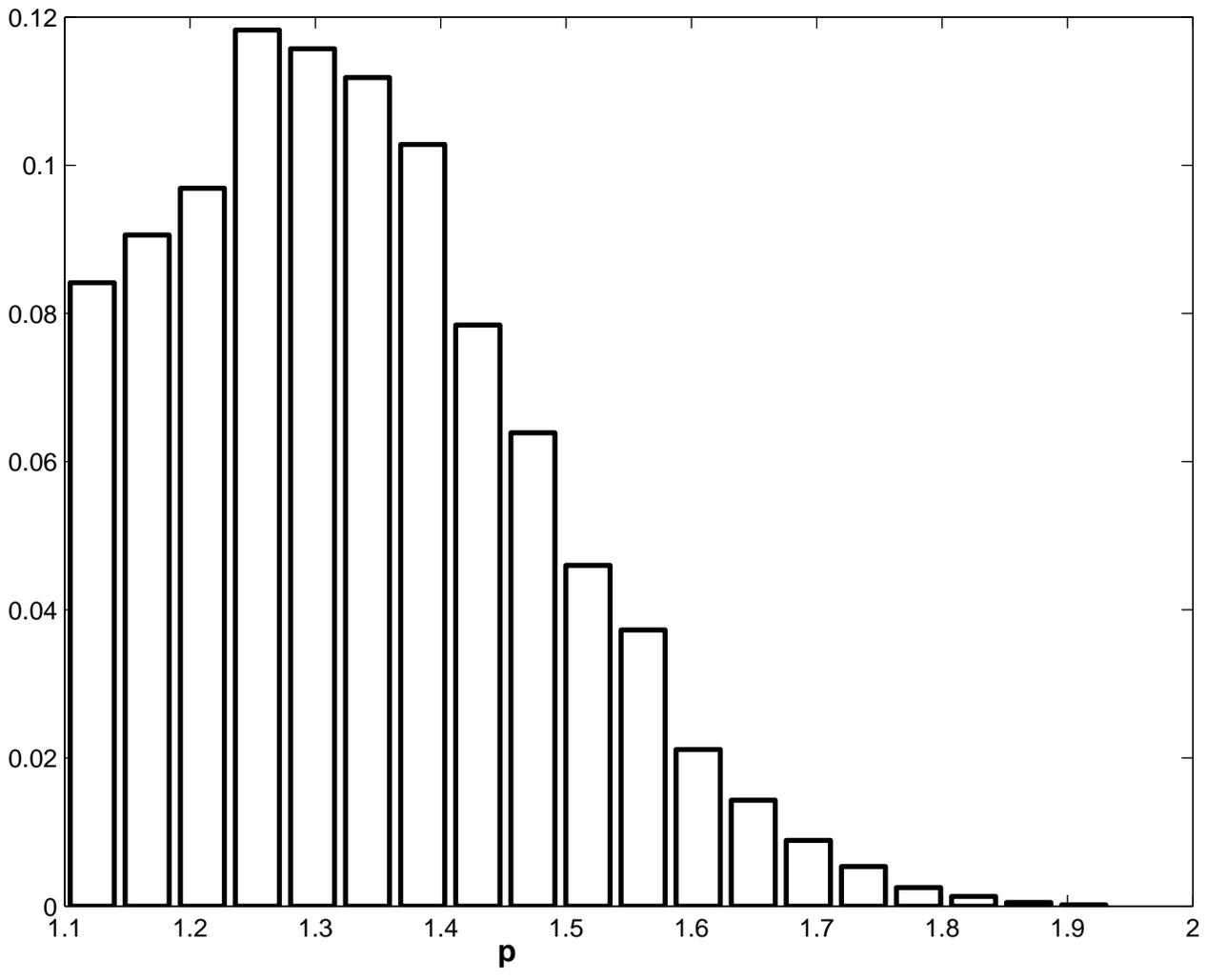}}\caption{Posterior
distribution of the model parameter $p$.} \label{fig7}
\end{figure}

As part of the model uncertainty analysis, it is useful to present plots of the
relevant quantities in the model selection (selection of $p$)
settings, see Figure \ref{fig6}, where we present
$ER_{p}=E[\widetilde {R}|\mathcal{D}_{I},p]$,
$PV_{p}=\sum_{i+j>I}E[\phi\left(\alpha_{i}\beta_{j}\right)
^{p}|\mathcal{D}_{I},p]$ and $EE_{p}=\mathrm{Var}(\widetilde
{R}|\mathcal{D}_{I},p)$ as a function of $p$. Figure \ref{fig6}
shows:

\begin{itemize}
\item MLE of $ER_{p}$ is almost constant, varying approximately from a maximum of
$603.96$ $(p=1.1)$ to a minimum of $595.78$ $(p=1.9)$ while the
MLE for $ER$ was 602.63.

\item The Bayesian estimates for $ER_{p}$ change as a function of $p.$ Approximately, it ranged
from a maximum of $646.4$ $(p=1.9)$ to a minimum of $621.1$
$(p=1.5)$ while the Bayesian estimator for $ER$ was $624.1$.
Hence, the difference (estimation bias) within this possible model
range is $\approx 25$ which is of a similar order as the process
uncertainty and the estimation error.

\item Bayesian estimators for $\sqrt{PV_{p}}$ and $\sqrt{EE_{p}}$ increase as $p$
increases approximately from $33.1$ to $68.5$ and from $37.4$ to
$102.0$ respectively, while the Bayesian estimators for
$\sqrt{PV}$ and $\sqrt{EE}$ are $37.3$ and $44.8$ correspondingly.
Hence, the resulting risk measure strongly varies in $p$ which has
a large influence on quantitative solvency requirements. The MLEs
for $PV_p$ and $EE_p$ are significantly less than the
corresponding Bayesian estimators. Also, the difference between
the MLE and the Bayesian estimators increases as $p$ increases.
\end{itemize}

For interpretation purposes of the above results it is helpful to
use the following relations between model averaging and model
selection quantities (easily derived from their definitions in Table
\ref{tab5}):
\begin{align}
ER &  =E[ER_p|\mathcal{D}_{I}], \\
PV & = E[PV_p|\mathcal{D}_{I}], \\
EE & = E[EE_p|\mathcal{D}_{I}]+\mathrm{Var}(ER_p|\mathcal{D}_{I}).
\end{align}
Here, the expectations are calculated with respect to the
posterior distribution of $p$. The histogram estimate of the later
is presented in Figure \ref{fig7} and highlights significant
uncertainty in $p$ (model uncertainty within Tweedie's compound
Poisson model).

\begin{figure}[ptb]
\centerline{\includegraphics{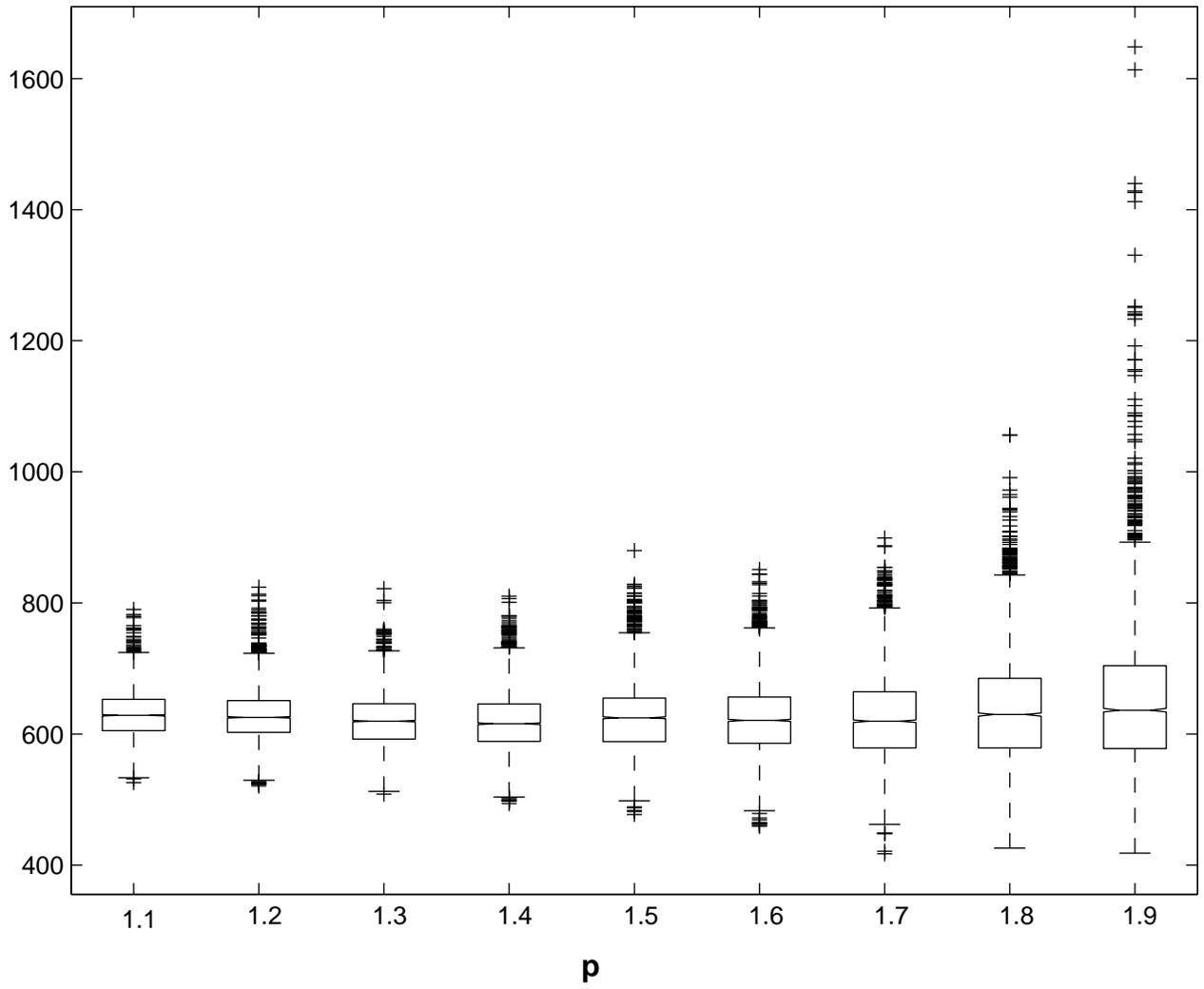}}\caption{Predicted
claim reserves $\widetilde{R}$ distributional summaries conditional
on model parameter $p$.} \label{fig8}
\end{figure}

We also provide Figure \ref{fig8} demonstrating a Box and Whisker
summary of the distributions of $\widetilde{R}~|~p$ for a range of
values of $p.\ $This plot provides the first, second and third
quartiles as the box. The notch represents uncertainty in the median
estimate for model comparison, across values of $p,$ and the
whiskers demonstrate the smallest and largest data points not
considered as outliers. The outliers are included as crosses and the
decision rule to determine if a point is an outlier was taken as the
default procedure from the statistical software package R.

The conclusion from this section is that if model selection is
performed (i.e.~$p$ is fixed by the modeller), the conditional
MSEP will increase significantly if a poor choice of the model
parameter $p$ is made. In particular, though the median is fairly
constant for the entire range of $p\in\left( 1,2\right)  $ the
shape of the distribution of $\widetilde{R}~|~p$ is clearly
becoming more diffuse as $p\rightarrow2$. This will lead to
significantly larger variance in the reserve estimate. If risk
measures such as Value-at-Risk are used in place of the mean, it
will result in reserves which are too conservative (if a poor
choice of $p$ is made). Also, using the maximum likelihood
approach may significantly underestimate the claims reserves and
associated uncertainties.

\subsection{Overdispersed Poisson and Gamma models}
There are several popular claims reserving models, however we
restrict our comparison to the overdispersed Poisson and gamma
models since they fit into Tweedie's compound Poisson framework when
$p=1$ and $p=2$ respectively. Note that the overdispersed Poisson
model and several other stochastic models lead to the same reserves
as the chain ladder method but different in higher moments. The
detailed treatment of these models can be found in e.g. England and
Verrall (2002) or W\"uthrich and Merz (2008), section 3.2.

The MLEs for the reserves and associated uncertainties within the
overdispersed Poisson and gamma models are provided in Table
\ref{tab8}. These results are obtained when the dispersion $\phi$ is
estimated by $\widehat{\phi}^{P}$ using Pearson's residuals
(\ref{DispersionViaPearson}) and when $\phi$ is estimated by
$\widehat{\phi}^{\mathrm{MLE}}$ obtained from the maximization of
the likelihood. The results for the first case are also presented in
W\"uthrich and Merz (2008), Table 6.4. Firstly note that, the values
of $\widehat{\phi}^{\mathrm{P}}$ and $\widehat{\phi}^{\mathrm{MLE}}$
are significantly different both for the overdispersed Poisson and
gamma models. As we mentioned in section 3.2, for a fixed $p$, the
MLE for the reserves does not depend on $\phi$ while the estimation
error, process variance and MSEP are proportional to $\phi$. As one
can see from Table \ref{tab8}, different estimators for the
dispersion $\phi$ lead to the same estimators for the reserves but
very different estimators for the uncertainties. Also note that, our
MLE calculations for Tweedie's distribution conditional on $p$, i.e.
Figure \ref{fig6}, are obtained using
$\widehat{\phi}^{\mathrm{MLE}}$ and are consistent with the
corresponding results for the overdispersed Poisson and Gamma models
when $p\rightarrow 1$ and $p\rightarrow 2$ respectively. Though, in
the case of the overdispersed Poisson we had to use an extended
quasi-likelihood to estimate $\widehat{\phi}^{\mathrm{MLE}}$. In
Figure \ref{fig6}, we do not show the results based on
$\widehat{\phi}^{P}$ but would like to mention that these are always
above the MLEs and below the Bayesian estimators for the process
variance and estimation error and are consistent with corresponding
overdispersed Poisson and gamma model limits. Interestingly, the
ratio $\widehat{\phi}^{P}/\widehat{\phi}^{\mathrm{MLE}}$ is
approximately $1.4-1.5$ for all considered cases of $p$ within a
range$[1,2]$.

The MLEs obtained using both $\widehat{\phi}^{\mathrm{MLE}}$ and
$\widehat{\phi}^{\mathrm{P}}$ underestimate the uncertainties
compared to the Bayesian analysis. Note that, while the MLEs for the
uncertainties are proportional to the dispersion estimator, the
corresponding Bayesian estimators are averages over all possible
values of $\phi$ according to its posterior distribution. The
uncertainty in the estimate for the dispersion is large which is
also highlighted by a bootstrap analysis in W\"uthrich and Merz
(2008), section 7.3. This indicates that $\phi$ should also depend
on the individual cells $(i,j)$. However, in this case
overparameterization needs to be considered with care and Bayesian
framework should be preferred.

\section{Discussion}

The results demonstrate the development of a Bayesian model for the
claims reserving problem when considering Tweedie's compound Poisson
model. The sampling methodology of a Gibbs sampler is applied to the
problem to study the model sensitivity for a real data set. The
problem of variable selection is addressed in a manner commensurate
with the MCMC sampling procedure developed in this paper and the
most probable model under the posterior marginal model probability
is then considered in further analysis. Under this model we then
consider two aspects, model selection and model averaging with
respect to model parameter $p$. The outcomes from these comparisons
demonstrate that the model uncertainty due to fixing $p$ plays a
significant role in the evaluation of the claims reserves and its
conditional MSEP. It is clear that whilst the frequentist MLE
approach is not sensitive to a poor model selection, the Bayesian
estimates demonstrate more dependence on poor model choice, with
respect to model parameter $p$. We use constant priors with very
wide ranges to perform inference in the setting where the posterior
is largely implied by data only. Also, we run a large number of MCMC
iterations so that numerical error in the Bayesian estimators is
very small. In the case of the data we studied, the MLEs for the
claims reserve, process variance and estimation error were all
significantly different (less) than corresponding Bayesian
estimators. This is due to the fact that the posterior distribution
implied by the data and estimated using MCMC is materially different
from Gaussian, i.e. more skewed.

Future research will examine variable selection aspects of this
model in a Bayesian context considering the entire set of possible
parameterizations. This requires development of advanced approaches
such as Reversible Jump MCMC and variable selection stochastic
optimization methodology to determine if a more parsimonious model
can be selected under assumptions of homogeneity in adjacent
columns/rows in the claims triangle.

\vspace{0.3cm} \noindent
\textbf{\large{Acknowledgements}}\\
\noindent The first author is thankful to the Department of
Mathematics and Statistics at the University of NSW for support
through an Australian Postgraduate Award and to CSIRO for support
through a postgraduate research top up scholarship. Thank you also
goes to Robert Kohn for discussions.

{\footnotesize{ Atchade Y. and Rosenthal, J. (2005) On adaptive
Markov chain Monte Carlo algorithms. \textit{Bernoulli}
\textbf{11}(5), 815-828.

\vspace{0.3cm} Bedard M. and Rosenthal\ J.S. (2008) Optimal scaling
of Metropolis algorithms: heading towards general target
distributions. \textit{The Canadian Journal of Statistics
}\textbf{36}(4), 483-503.

\vspace{0.3cm} Bernardo, J.M. and Smith, A.F.M. (1994)
\textit{Bayesian Theory}. John Wiley and Sons, NY.

\vspace{0.3cm} Cairns, A.J.G. (2000) A discussion of parameter and
model uncertainty in insurance.\textit{ Insurance: Mathematics and
Economics }\textbf{27}, 313-330.

\vspace{0.3cm} Carlin, B. and Chib, S. (1995) Bayesian model choice
via Markov chain Monte Carlo methods.\textit{ Journal of the Royal
Statististical Society Series B} \textbf{57}, 473-484.

\vspace{0.3cm} Casella, G. and George, E.I. (1992) Explaining the
Gibbs Sampler. \textit{The American Statistician} \textbf{46}(3),
167-174.

\vspace{0.3cm} Congdon P. (2006)\textit{ }Bayesian model choice
based on Monte Carlo estimates of posterior model probabilities.
\textit{Computational Statistics and Data Analysis} \textbf{50}(2),
346-357.

\vspace{0.3cm} Dunn, P.K. and Smyth, G.K. (2005) Series evaluation
of Tweedie exponential dispersion model densities.\textit{
Statistics and Computing} \textbf{15}, 267-280.

\vspace{0.3cm} England P.D. and Verrall R.J. (2002) Stochastic
claims reserving in general insurance.\textit{ British Actuarial
Journal} \textbf{8}(3), 443-510.

\vspace{0.3cm} Gelman, A., Carlin, J.B., Stern, H.S. and Rubin, D.B.
(1995) \textit{ Bayesian Data Analysis}. Chapman and Hall /CRC Texts
in Statistical Science Series, \textbf{60}.

\vspace{0.3cm} Gelman, A., Gilks, W.R. and Roberts, G.O. (1997) Weak
convergence and optimal scaling of random walks metropolis
algorithm. \textit{Annals of~Applied~Probability} \textbf{7},
110-120.

\vspace{0.3cm} Gilks, W.R., Richardson, S. and Spiegelhalter, D.J.
(1996) \textit{Markov Chain Monte Carlo in Practice}. Chapman and
Hall, Florida.

\vspace{0.3cm} Green, P. (1995) Reversible jump Markov chain Monte
Carlo computation and Bayesian model determination.
\textit{Biometrika} \textbf{82}, 711-732.

\vspace{0.3cm} J{\o}rgensen, B. and de Souza, M.C.P. (1994) Fitting
Tweedie's compound Poisson model to insurance claims data.
\textit{Scandinavian Actuarial Journal}, 69-93.

\vspace{0.3cm} Robert, C.P. and\ Casella, G. (2004) \textit{Monte
Carlo Statistical Methods}, 2nd Edition Springer Texts in
Statistics.

\vspace{0.3cm} Roberts,\ G.O. and Rosenthal, J.S. (2001) Optimal
scaling for various Metropolis-Hastings algorithms.
\textit{Statistical Science} \textbf{16}, 351-367.

\vspace{0.3cm} Rosenthal, J.S. (2007) AMCMC: An R interface for
adaptive MCMC. \textit{Computational Statistics and Data\ Analysis}
\textbf{51}(12), 5467-5470.

\vspace{0.3cm} Smith, A.F.M. and Roberts, G.O. (1993) Bayesian
computation via the Gibbs sampler and related Markov chain Monte
Carlo methods. \textit{Journal of Royal Statistical Society Series
B} \textbf{55}(1), 3-23.

\vspace{0.3cm} Smyth, G.K. and J{\o}rgensen, B. (2002) Fitting
Tweedie's compound Poisson model to insurance claims data:
dispersion modelling. \textit{Astin Bulletin} \textbf{32}, 143-157.

\vspace{0.3cm} Tweedie, M.C.K. (1984) An index which distinguishes
between some important exponential families. In Statistics:
Applications in new directions. \textit{Proceeding of the Indian
Statistical Institute Golden Jubilee International Conference},
J.K.~Ghosh and J.~Roy (eds.), 579-604, Indian Statistical Institute
Canada.

\vspace{0.3cm} Wright, E.M. (1935) On asymptotic expansions of
generalized Bessel functions. \textit{Proceedings of London
Mathematical Society} \textbf{38}, 257-270.

\vspace{0.3cm} W\"{u}thrich, M.V. (2003) Claims reserving using
Tweedie's compound Poisson model. \textit{Astin Bulletin}
\textbf{33}, 331-346.

\vspace{0.3cm} W\"{u}thrich, M.V. and Merz, M. (2008)
\textit{Stochastic Claims Reserving Methods in Insurance,} Wiley
Finance. } }

\vspace{2.0cm}

{\bf \large{Gareth W.~Peters} }\\
\textit{CSIRO Mathematical and Information Sciences, Sydney, Locked
Bag 17, North Ryde, NSW, 1670, Australia}\\
and \\
\textit{UNSW Mathematics and Statistics Department, Sydney, 2052,
Australia. \\
Email: peterga@maths.unsw.edu.au }\\

{\bf \large{Pavel V.~Shevchenko}} (Corresponding Author)\\
\textit{CSIRO Mathematical and Information Sciences, Sydney, Locked
Bag 17, North Ryde, NSW, 1670, Australia.\\
Email: Pavel.Shevchenko@csiro.au }\\

{\bf \large{Mario V.~W\"{u}thrich}} \\
\textit{ETH Zurich, Department of Mathematics, CH-8092 Zurich, Switzerland. \\
Email: wueth@math.ethz.ch}\\

\newpage

\begin{table}[ptb]
\begin{center}
{\footnotesize
\begin{tabular}
[c]{|c||rrrrrrrrrr|}\hline
accident & \multicolumn{10}{c|}{development years $j$}\\
year $i$ & \quad0 \quad & \quad1 \quad & \quad\quad & \quad\quad & \quad\quad
& \enspace\dots\quad & \quad$j$\quad & \quad~ \quad & \enspace\dots\quad &
\quad$I$ \quad\\\hline
$0$ & \multicolumn{10}{c|}{}\\\cline{11-11}%
$1$ & \multicolumn{9}{c|}{observed random variables $Y_{i,j}\in\mathcal{D}%
_{I}$} & \multicolumn{1}{c|}{}\\\cline{10-10}%
$\vdots$ & \multicolumn{8}{c|}{} &
\multicolumn{2}{c|}{}\\\cline{9-9} & \multicolumn{7}{c|}{} &
\multicolumn{3}{c|}{}\\\cline{8-8} $i$ & \multicolumn{6}{c|}{} &
\multicolumn{4}{c|}{}\\\cline{7-7} & \multicolumn{5}{c|}{} &
\multicolumn{5}{c|}{}\\\cline{6-6} & \multicolumn{4}{c|}{} &
\multicolumn{6}{c|}{}\\\cline{5-5} $\vdots$ & \multicolumn{3}{c|}{}
& \multicolumn{7}{c|}{to be predicted
${Y}_{i,j}\in\mathcal{D}_{I}^{c}$}\\\cline{4-4}%
$I-1$ & \multicolumn{2}{c|}{} & \multicolumn{8}{c|}{}\\\cline{3-3}
$I$ & \multicolumn{1}{c|}{} & \multicolumn{9}{c|}{}\\\hline
\end{tabular}
}
\end{center}
\caption{Claims development triangle.}
\label{tab1}
\end{table}

\begin{table}[ptbh]
\begin{center}
{\scriptsize {\
\begin{tabular}
[c]{c|cccccccccc}
{Year} & {$0$} & {$1$} & {$2$} & {$3$} & {$4$} &
{$5$} & {$6$} & {$7$} & {$8$} & {$9$}\\\hline {$0$} & {$594.6975$} &
{$372.1236$} & {$89.5717$} & {$20.7760$} & {$20.6704 $} & {$6.2124$}
& {$6.5813$} & {$1.4850$} & {$1.1130$} &
\multicolumn{1}{c|}{$1.5813$}\\\cline{11-11} {$1$} & {$634.6756$} &
{$324.6406$} & {$72.3222$} & {$15.1797$} & {$6.7824$} & {$3.6603$} &
{$5.2752$} & {$1.1186$} & {$1.1646$} & \multicolumn{1}{|c|}{}
\\\cline{10-10}
{$2$} & {$626.9090$} & {$297.6223$} & {$84.7053$} & {$26.2768$} &
{$15.2703 $} & {$6.5444$} & {$5.3545$} & {$0.8924$} &
\multicolumn{1}{|c}{} & \multicolumn{1}{c|}{}\\\cline{9-9} {$3$} &
{$586.3015$} & {$268.3224$} & {$72.2532$} & {$19.0653$} & {$13.2976
$} & {$8.8340$} & {$4.3329$} & \multicolumn{1}{|c}{} &  &
\multicolumn{1}{c|}{}\\\cline{8-8} {$4$} & {$577.8885$} &
{$274.5229$} & {$65.3894$} & {$27.3395$} & {$23.0288
$} & {$10.5224$} & \multicolumn{1}{|c}{} &  &  & \multicolumn{1}{c|}{}%
\\\cline{7-7}
{$5$} & {$618.4793$} & {$282.8338$} & {$57.2765$} & {\ $24.4899$} &
{$10.4957$} & \multicolumn{1}{|c}{} &  &  &  & \multicolumn{1}{c|}{}%
\\\cline{6-6}%
{$6$} & {$560.0184$} & {$289.3207$} & {$56.3114$} & {\ $22.5517$} &
\multicolumn{1}{|c}{} &  &  &  &  &
\multicolumn{1}{c|}{}\\\cline{5-5} {$7$} & {$528.8066$} &
{$244.0103$} & {$52.8043$} & \multicolumn{1}{|c}{} & &  &  &  &  &
\multicolumn{1}{c|}{}\\\cline{4-4} {$8$} & {$529.0793$} &
{$235.7936$} & \multicolumn{1}{|c}{} &  &  &  &  &  & &
\multicolumn{1}{c|}{}\\\cline{3-3} {$9$} & {$567.5568$} &
\multicolumn{1}{|c}{} &  &  &  &  &  &  &  &
\multicolumn{1}{c|}{}\\\cline{2-11}
\end{tabular}
} }
\end{center}
\caption{Data - annual claims payments $Y_{i,j}$ for each accident year $i$
and development year $j,$ $i+j\leq9$.}
\label{tab2}
\end{table}

\newpage

\begin{table}[ptbh]
\begin{center}
{\footnotesize{
\begin{tabular}
[c]{p{0.2in}|p{30pt}|c|p{70pt}|p{72pt}|p{72pt}|c|}\cline{2-7}%
\raisebox{-1.50ex}[0cm][0cm]{} & \textbf{MLE} & \textbf{MLE stdev} &
\multicolumn{3}{|p{225pt}|}{\textbf{Bayesian posterior}} &
$\sigma_{RW} $\\\cline{4-6} &  &  & \textbf{MMSE} & \textbf{stdev} &
$[Q_{0.05};Q_{0.95}]$ & \\\hline \multicolumn{1}{|p{0.2in}|}{$p$} &
1.259 & 0.149 & 1.332 (0.007) & 0.143 (0.004) & [1.127;1.590] &
1.61\\\hline \multicolumn{1}{|p{0.2in}|}{$\phi$} & 0.351 & 0.201 &
0.533 (0.013) & 0.289 (0.005) & [0.174;1.119] & 1.94\\\hline
\multicolumn{1}{|p{0.2in}|}{$\alpha_{1}$} & 0.918 & 0.056 & 0.901
(0.004) & 0.074 (0.001) & [0.778;1.022] & 0.842\\\hline
\multicolumn{1}{|p{0.2in}|}{$\alpha_{2}$} & 0.946 & 0.051 & 0.946
(0.003) & 0.073 (0.001) & [0.833;1.072] & 0.907\\\hline
\multicolumn{1}{|p{0.2in}|}{$\alpha_{3}$} & 0.861 & 0.048 & 0.861
(0.003) & 0.068 (0.001) & [0.756;0.977] & 0.849\\\hline
\multicolumn{1}{|p{0.2in}|}{$\alpha_{4}$} & 0.891 & 0.049 & 0.902
(0.003) & 0.072 (0.002) & [0.794;1.027] & 0.893\\\hline
\multicolumn{1}{|p{0.2in}|}{$\alpha_{5}$} & 0.879 & 0.051 & 0.876
(0.003) & 0.070 (0.001) & [0.768;0.994] & 0.932\\\hline
\multicolumn{1}{|p{0.2in}|}{$\alpha_{6}$} & 0.842 & 0.048 & 0.843
(0.002) & 0.069 (0.001) & [0.736;0.958] & 0.751\\\hline
\multicolumn{1}{|p{0.2in}|}{$\alpha_{7}$} & 0.762 & 0.046 & 0.762
(0.003) & 0.066 (0.001) & [0.660;0.876] & 0.888\\\hline
\multicolumn{1}{|p{0.2in}|}{$\alpha_{8}$} & 0.763 & 0.047 & 0.765
(0.003) & 0.067 (0.001) & [0.661;0.874] & 0.897\\\hline
\multicolumn{1}{|p{0.2in}|}{$\alpha_{9}$} & 0.848 & 0.059 & 0.856
(0.003) & 0.090 (0.002) & [0.716;1.009] & 1.276\\\hline
\multicolumn{1}{|p{0.2in}|}{$\beta_{0}$} & 669.1 & 27.7 & 672.7
(2.1) & 39.7 (0.7) & [610.0;740.0] & 296\\\hline
\multicolumn{1}{|p{0.2in}|}{$\beta_{1}$} & 329.0 & 14.4 & 331.1
(1.0) & 20.6 (0.4) & [298.1;365.9] & 190\\\hline
\multicolumn{1}{|p{0.2in}|}{$\beta_{2}$} & 77.43 & 4.38 & 78.06
(0.24) & 6.10 (0.06) & [68.58;88.29] & 75.4\\\hline
\multicolumn{1}{|p{0.2in}|}{$\beta_{3}$} & 24.59 & 1.96 & 24.95
(0.08) & 2.64 (0.03) & [20.89;29.64] & 40.9\\\hline
\multicolumn{1}{|p{0.2in}|}{$\beta_{4}$} & 16.28 & 1.55 & 16.65
(0.05) & 2.09 (0.03) & [13.44;20.30] & 40.6\\\hline
\multicolumn{1}{|p{0.2in}|}{$\beta_{5}$} & 7.773 & 1.028 & 8.068
(0.024) & 1.356 (0.020) & [6.064;10.473] & 26.0\\\hline
\multicolumn{1}{|p{0.2in}|}{$\beta_{6}$} & 5.776 & 0.937 & 6.115
(0.022) & 1.261 (0.016) & [4.246;8.347] & 24.1\\\hline
\multicolumn{1}{|p{0.2in}|}{$\beta_{7}$} & 1.219 & 0.396 & 1.494
(0.006) & 0.609 (0.013) & [0.739;2.609] & 13.1\\\hline
\multicolumn{1}{|p{0.2in}|}{$\beta_{8}$} & 1.188 & 0.476 & 1.622
(0.008) & 0.802 (0.016) & [0.674;3.070] & 15.1\\\hline
\multicolumn{1}{|p{0.2in}|}{$\beta_{9}$} & 1.581 & 0.790 & 2.439
(0.021) & 1.496 (0.026) & [0.829;5.250] & 32.1\\\hline
\end{tabular}
}}
\end{center}
\caption{MLE and Bayesian estimators. $\sigma_{RW}$ is the proposal
standard deviation in the MCMC algorithm and $\left[
Q_{0.05};Q_{0.95}\right] $ is the predictive interval, where
$Q_{\alpha}$ is the quantile of the posterior distribution at level
$\alpha$. The numerical standard error, in Bayesian estimators due
to finite number of MCMC iterations, is included in brackets next to
estimates.} \label{tab3}
\end{table}

\newpage

\begin{table}[ptb]
\begin{center}
{\footnotesize {\
\begin{tabular}
[c]{|c|ccccccc|}\hline & $M_{0}$ & $M_{1}~$ & $M_{2}~$ & $M_{3}~$ &
$M_{4}~$ & $M_{5}~$ & $M_{6} ~$\\\hline 
$\pi(M_{k}~|~D_{I})$
& 0.71 & 4.19E-54 & 3.04E-43 & 1.03E-28 & 6.71E-20 & 2.17E-21 &
0.29\\\hline DIC & 399 & 649 & 600 & 535 & 498 & 507 & 398\\\hline
LHR $p-value$ & 1 & 2.76E-50 & 1.67E-40 & 3.53E-28 & 5.78E-21 &
3.03E-23 & 0.043\\\hline
\end{tabular}
}}
\end{center}
\caption{Posterior model probabilities $\pi\left(
M_{k}|D_{I}\right)$, Deviance Information Criterion (DIC) for
variable selection models $M_0,\ldots,M_6$ and Likelihood Ratio (LHR)
p-values (comparing $M_0$ to $M_1,\ldots,M_6$).} \label{tab4}
\end{table}

\begin{table}[ptb]
\begin{center}
{\footnotesize{
\begin{tabular}
[c]{c|c|c|}\cline{2-3} & \textbf{Model Averaging} & \textbf{Model
Selection for} $p$\\\hline 
\multicolumn{1}{|c|}{Estimated Reserves} &
$ER=\widehat{R}^{\mathrm{B} }=E[ \widetilde{R}|\mathcal{D}_{I}] $ &
$ER_{p}=E[ \widetilde{R}| \mathcal{D}_{I},p]
$\\
\multicolumn{1}{|c|}{Process Variance} & $PV=E\left[  \left.  \sum\phi\left(
\alpha_{i}\beta_{j}\right)  ^{p}\right\vert \mathcal{D}_{I}\right]  $ &
$PV_{p}=E\left[  \left.  \sum\phi\left(  \alpha_{i}\beta_{j}\right)
^{p}\right\vert \mathcal{D}_{I},p\right]  $\\
\multicolumn{1}{|c|}{Estimation Error} &
$EE=\mathrm{Var}(\widetilde{R}| \mathcal{D}_{I})  $ &
$EE_{p}=\mathrm{Var} ( \widetilde{R}|\mathcal{D}_{I},p) $\\\hline
\end{tabular}
}}
\end{center}
\caption{Quantities used for analysis of the claims reserving
problem under Model Averaging and Model Selection in respect to
$p$.}
\label{tab5}
\end{table}

\begin{table}[ptb]
\begin{center}
{\footnotesize{
\begin{tabular}
[c]{c|c|c|}\cline{2-3} & \multicolumn{2}{|c|}{\textbf{Model
Averaging}}\\\hline \multicolumn{1}{|c|}{{Statistic}} & {Bayesian
Estimate} & MLE Estimate\\\hline
\multicolumn{1}{|c|}{$ER$} & 624.1 (0.7) & 602.630\\
\multicolumn{1}{|c|}{$\sqrt{PV}$} & 37.3 (0.2) & 25.937\\
\multicolumn{1}{|c|}{$\sqrt{EE}$} & 44.8 (0.5) & 28.336\\
\multicolumn{1}{|c|}{$\sqrt{MSEP}$} & 58.3(0.5) & 38.414\\ \hline
\end{tabular}
}}
\end{center}
\caption{Model averaged estimates of claim reserve, process variance
and estimation error. Numerical error in Bayesian estimates is
reported in brackets. See Table \ref{tab5} for definitions of ER,
PV, EE and MSEP=EE+PV.} \label{tab6}
\end{table}

\begin{table}[ptb]
\begin{center}
{\footnotesize{
\begin{tabular}
[c]{c|c|c|}\cline{2-3} & \multicolumn{2}{|c|}{\textbf{Model
Averaging}}\\\hline \multicolumn{1}{|c|}{{{VaR}$_{q}$}} & {${R}$} &
$\widetilde{R}$\\\hline
\multicolumn{1}{|c|}{VaR$_{75\%}$} & 659.8 (0.9) & 650.6 (1.0)\\
\multicolumn{1}{|c|}{VaR$_{90\%}$} & 698.4 (1.2) & 680.4 (1.3)\\
\multicolumn{1}{|c|}{VaR$_{95\%}$} & 724.0 (1.5) & 701.7
(1.6)\\\hline
\end{tabular}
}}
\end{center}
\caption{Bayesian model averaged estimates of Value at Risk for
outstanding claims payment $R$ and claim reserves $\widetilde{R}$.}
\label{tab7}
\end{table}

\begin{table}[ptb]
\begin{center}
{\footnotesize{
\begin{tabular}
[c]{c|c|c|c|c|}\cline{2-5} &
\multicolumn{2}{|c|}{\textbf{Overdispersed Poisson}} &
\multicolumn{2}{|c|} {\textbf{Gamma model}}
\\ \cline{1-5} \multicolumn{1}{|c|}{Statistic} &
$\widehat{\phi}^{P}\approx 1.471$ &
$\widehat{\phi}^{MLE}\approx 0.954$ & $\widehat{\phi}^{P}\approx 0.045$ & $\widehat{\phi}^{MLE}\approx 0.031$ \\
\hline 
\multicolumn{1}{|c|}{$ER_p$} & 604.706 & 604.706 & 594.705 & 594.705 \\
\multicolumn{1}{|c|}{$\sqrt{PV_p}$} & 29.829 &  24.017 & 62.481 & 52.162 \\
\multicolumn{1}{|c|}{$\sqrt{EE_p}$} & 30.956 & 24.925 & 92.826 & 77.496 \\
\multicolumn{1}{|c|}{$\sqrt{MSEP_p}$} & 42.989 & 34.613 & 111.895 & 93.415\\
\hline
\end{tabular}
}}
\end{center}
\caption{The MLEs for the overdispersed Poisson ($p=1$) and Gamma
($p=2$) models, when the dispersion $\phi$ is estimated as
$\widehat{\phi}^{\mathrm{P}}$ using Pearson's residuals
(\ref{DispersionViaPearson}) or $\widehat{\phi}^{\mathrm{MLE}}$.}
\label{tab8}
\end{table}

\newpage

\newpage

\newpage

\newpage

\newpage

\newpage

\newpage

\newpage

\newpage
\end{document}